\documentclass[12pt,preprint]{aastex}
\usepackage{graphicx}

\usepackage{draftwatermark}
\SetWatermarkText{Pre-print}

\def\mb{\ifmmode {{\rm B_{435}}}\else
                ${\rm B_{435}}$\fi}
\def\mv{\ifmmode {{\rm V_{606}}}\else
                ${\rm V_{606}}$\fi}
\def\mi{\ifmmode {{\rm i_{775}}}\else
                ${\rm i_{775}}$\fi}
\def\mz{\ifmmode {{\rm z_{850}}}\else
                ${\rm z_{850}}$\fi}

\def\mY{\ifmmode {{\rm Y_{105}}}\else
                ${\rm Y_{105}}$\fi}
\def\mJ{\ifmmode {{\rm J_{125}}}\else
                ${\rm J_{125}}$\fi}
\def\mH{\ifmmode {{\rm H_{160}}}\else
                ${\rm H_{160}}$\fi}
            
\def\Lya{Ly-$\alpha$}
\def\OIII{[OIII]}
\def\Ha{H$\alpha$}

\def\Hb{H$\beta$}
\def\Hg{H$\gamma$}
\def\OII{[OII]}
\def\CIII{CIII]}
\def\CIV{CIV}
\def\NeIII{[NeIII]}
\def\MgII{MgII}

\def\Gini{$G$}
\def\M20{$M_{20}$}
\def\vmax{${\rm 1/V_{max}}\ $}

\begin{document}
\title{Emission-Line Galaxies from the Hubble Space Telescope Probing Evolution and Reionization Spectroscopically (PEARS) Grism Survey. II:  The Complete Sample}
\author{Nor Pirzkal\altaffilmark{1}, Barry Rothberg\altaffilmark{1,2,3}\\ Chun Ly\altaffilmark{1,4}, Sangeeta Malhotra\altaffilmark{5}, James E. Rhoads\altaffilmark{5}, Norman A. Grogin\altaffilmark{1}, Tomas Dahlen\altaffilmark{1}, Kai G. Noeske \altaffilmark{1}, Gerhardt R. Meurer\altaffilmark{6}, Jeremy R. Walsh\altaffilmark{7}, Nimish P. Hathi\altaffilmark{8},  Seth H. Cohen\altaffilmark{5}, Andrea Bellini\altaffilmark{1}, Benne W. Holwerda\altaffilmark{9}, Amber N. Straughn\altaffilmark{10}, Matthew Mechtley\altaffilmark{5}, Rogier A. Windhorst \altaffilmark{5}
 }
\altaffiltext{1}{Space Telescope Science Institute, 3700 San Martin Drive, Baltimore, MD, 21210,USA}
\altaffiltext{2}{George Mason University, Department of Physics \& Astronomy, MS 3F3, 4400 University Drive, Fairfax, VA 22030, USA}
\altaffiltext{3}{Leibniz-Institut f\"{u}r Astrophysik Potsdam (AIP), An der Sternwarte 16, 14482, Potsdam, Germany}
\altaffiltext{4}{Giacconi Fellow}
\altaffiltext{5}{School of Earth And Space Exploration, Arizona State University, Tempe, AZ, 85287-1404, USA}
\altaffiltext{6}{International Centre for Radio Astronomy Research,The University of Western Australia,35 StirlingHighway, Crawley, WA 6009, Australia}
\altaffiltext{7}{European Southern Observatory, Karl-Schwarzschild-Strasse 2, D-85748 Garching, Germany}
\altaffiltext{8}{Carnegie Observatories, 813 Santa Barbara Street, Pasadena, CA 91101, USA}
\altaffiltext{9}{ESA Fellow, ESA-ESTEC, Keplerlaan 1, 2200 AG, Noordwijk, the Netherlands }
\altaffiltext{10}{Astrophysics Science Division, Goddard Space Flight Center, Code 665, Greenbelt, MD 20771, USA}

\begin{abstract}
We present a full analysis of the Probing Evolution And Reionization Spectroscopically
(PEARS) slitess grism spectroscopic data obtained with the Advanced Camera for Surveys
onboard {\it HST}. PEARS covers fields within both the Great Observatories Origins
Deep Survey (GOODS) North and South fields, making it ideal as a random survey
of galaxies, as well as the availability of a wide variety of ancillary observations
complemented by the spectroscopic results.  Using the PEARS data we are able to identify 
star-forming galaxies within the redshift volume 0 $<$ z $<$ 1.5.  Star-forming regions 
in the PEARS survey are pinpointed independently of the host galaxy.  This method allows 
us to detect the presence of multiple emission line regions (ELRs) within a single galaxy. 
We identified a total of 1162  \Ha, \OIII\ and/or \OII\ emission-lines in the PEARS 
sample of  906 galaxies to a limiting flux of 
$\sim$ ${\rm 10^{-18}\ erg/s/cm^2}$.   The ELRs have also been compared to the properties 
of the host galaxy, including morphology, luminosity, and mass.  From this analysis, we 
find three key results: 1)  The computed line luminosities show evidence of a flattening 
in the luminosity function with increasing redshift; 2) The star-forming systems show 
evidence of complex morphologies, with star formation occurring predominantly within 
one effective (half-light) radius. However, the morphologies show no correlation with 
host stellar mass;  and 3) The number density of star-forming galaxies with 
{\it M}$_{*}$ $\geq$ 10$^{9}$ {\it M}$_{\odot}$ decreases by an order of 
magnitude at z $\leq$ 0.5 relative to the number at 0.5 $<$ z $<$ 0.9 supporting
 the argument of galaxy downsizing.
\end{abstract}

\section{Introduction}
\indent Emission line galaxies (ELGs) are systems selected by the presence of strong 
line emissions (e.g., \Lya\, \OII, \OIII, \Hb, and \Ha), usually detected using narrow band  or grism
surveys.  The strong emission lines in these galaxies trace recent star formation activity, 
in contrast to the star formation history and properties of the global stellar populations that can be
discerned using broad band observations.  A significant amount of the light originating from ELGs 
is contained in regions producing strong emission lines, which in turns makes these objects easily identifiable.
The emission lines of ELGs also provide a convenient way to determine the redshifts of these objects.
Since ELGs are selected on the basis of strong emission lines, rather than continuum emission, selecting
 ELGs allows one to potentially probe down to a  lower luminosity --- and 
thus lower mass -- galaxies compared to broad band surveys, which tend
to be limited by the luminosity of the galaxies themselves, rather than the strength of their emission  lines.   Assuming 
that ELGs are spatially distributed in a fashion similar to other galaxies, they provide 
a powerful tool for tracing the star formation history of the Universe \citep[e.g.,][]{Salzer1988,Popescu1997}.\\

\indent The epoch $0 < z < 1.5$  discussed in this paper is important, because star formation activity in galaxies
has been observed to increase significantly  as redshift increases \citep[e.g.][]{Madau1998,hopkins2004}. While at higher redshifts ($z>2$), there is 
 still some controversy as to whether the star-formation rate density (SFRD) relation flattens or decreases, the
initial increase in star-formation  implies that, at low z, some mechanism(s) must have
occurred, which quenched star formation. If this was not the case, massive ellipticals today
would still be observed to be strongly forming many stars. There is also evidence that suggest that the interstellar medium, star formation
rates and gas fractions differ between local and distant galaxies. 
Studying galaxy evolution at these redshifts therefore does not only require the ability to measure the star formation history of these objects, 
but also depends on our ability to properly sample galaxies over a wide range of masses to alleviate as many biases as possible.  
ELGs are ideal for such work.  As noted above, these objects are easily detected in surveys and they
are efficient for probing to lower stellar masses in terms of telescope 
time required.  The wavelength range of the ACS grism used for PEARS makes it possible
to identify the strong rest-frame emission lines that are well known to be a sign of vigorous star formation (e.g. \Ha, \OIII\ and \OII) 
out to z $\simeq$ 1.5.  In this paper,  examining  \Ha, \OIII\ and \OII\  emitters separately allows us to 
look at properties of star-forming galaxies in increasing redshift ranges.  When plotted separately, these three emission lines represent
proxies for the redshift bins of $0<z<0.5$, $0.1<z<0.9$, and $0.5<z<1.5$, respectively.\\

\indent Identifying ELGs has traditionally been done using narrow-band photometric
filters.  This technique has also been successfully applied to very high redshifts to 
detect \Lya\ emitters \citep{Rhoads2001}. However, while narrow-band surveys can efficiently cover large fields-of-view 
 to relatively faint magnitudes, they are typically limited to very small
and discrete redshift ranges.  This can be partially alleviated using multiple narrow
band filters \citep[e.g. See Subaru Deep Field,][]{Ly2007}, but continuous redshift coverage remains
intrinsically limited in narrow-band surveys.
The Probing Evolution And Reionization Spectroscopically
(PEARS) slitless grism spectroscopic survey provides an unprecedented opportunity to study
ELGs in a way that cannot be achieved from any ground-based observations.  PEARS allows 
us to bypass the difficulties inherent in narrow-band filter surveys (as noted above)
and the limitations imposed by varying sky-brightness and atmospheric emission lines,
which can limit ground-based grism surveys, and identify ELGs based {\it solely} on the 
direct detection of emission lines in dispersed slitless spectra.
As previously shown \citep{pirzkal2006,straughn2008,straughn2009}, this approach allows us to detect 
emission lines in very faint host galaxies, particularly sub-{\it m}$^{*}$ galaxies, over a very large and 
continuous redshift range. Since our survey is mainly limited by the line fluxes themselves, faint emission lines can be identified in galaxies that 
are only weakly detected (i.e. high EW emission lines) while brighter host galaxies tend to increase the local background flux, diluting fainter emission lines in the brightest hosts. The line flux limit of
this survey is discussed in more details in Section \ref{completeness}.
Particular to the PEARS survey, the use of multiple position
angles on the sky allows us to identify emission lines using independent observations, as well as
 to pinpoint the exact physical location of the ELRs within each ELG.
Crucially, and in addition to this, the PEARS survey was designed to overlap with both the GOODS-N and GOODS-S fields, 
so that there exists a substantial amount of ancillary data available, including very deep, high-resolution broad-band imaging ranging from the UV to the infrared bands. 

As we noted above, the redshift range (0 $<$ z $<$ 1.5) probed by PEARS is a critical transition
epoch, both in terms of star-formation histories and morphological evolution. On one hand, the PEARS
grism slitless observations make it possible to efficiently identify emission lines, identify the corresponding ELRs and host
ELGs.  On the other hand, the GOODS ancillary data allow us to examine the morphology and physical
characteristics of the ELGs. This powerful combination of data gives us an opportunity to examine the evolution of ELGs over
 a long period of cosmic time and over a much wider mass range than has been previously probed.\\
 
\indent This paper is organized as follow: Section \ref{observations} briefly summarizes 
the PEARS observations (HST Proposal 10530, P.I. Malhotra). Section \ref{analysis} 
describes the data reduction and analysis of the sample, including detection, extraction 
and identification of emission lines, as well as completeness tests.
Section \ref{ELR} presents the PEARS  \OII, \OIII\ and \Ha\ line luminosity 
functions and their redshift evolution. Finally, Section \ref{hosts} compares the 
properties of the PEARS host galaxies, such as morphology and luminosity, with
the star formation properties discerned from the PEARS emission lines.
  All calculations in this paper assume $H_0={\rm 70\ km\ s^{-1}\ Mpc^{-1}}$ and $\Omega_M  = 0.3$, $\Omega_\Lambda = 0.7$ \citep{Komatsu2011,Hinshaw2012}. All magnitudes are given in the AB system.

\section{Observations}\label{observations}
\begin{figure}
\includegraphics[width =7in]{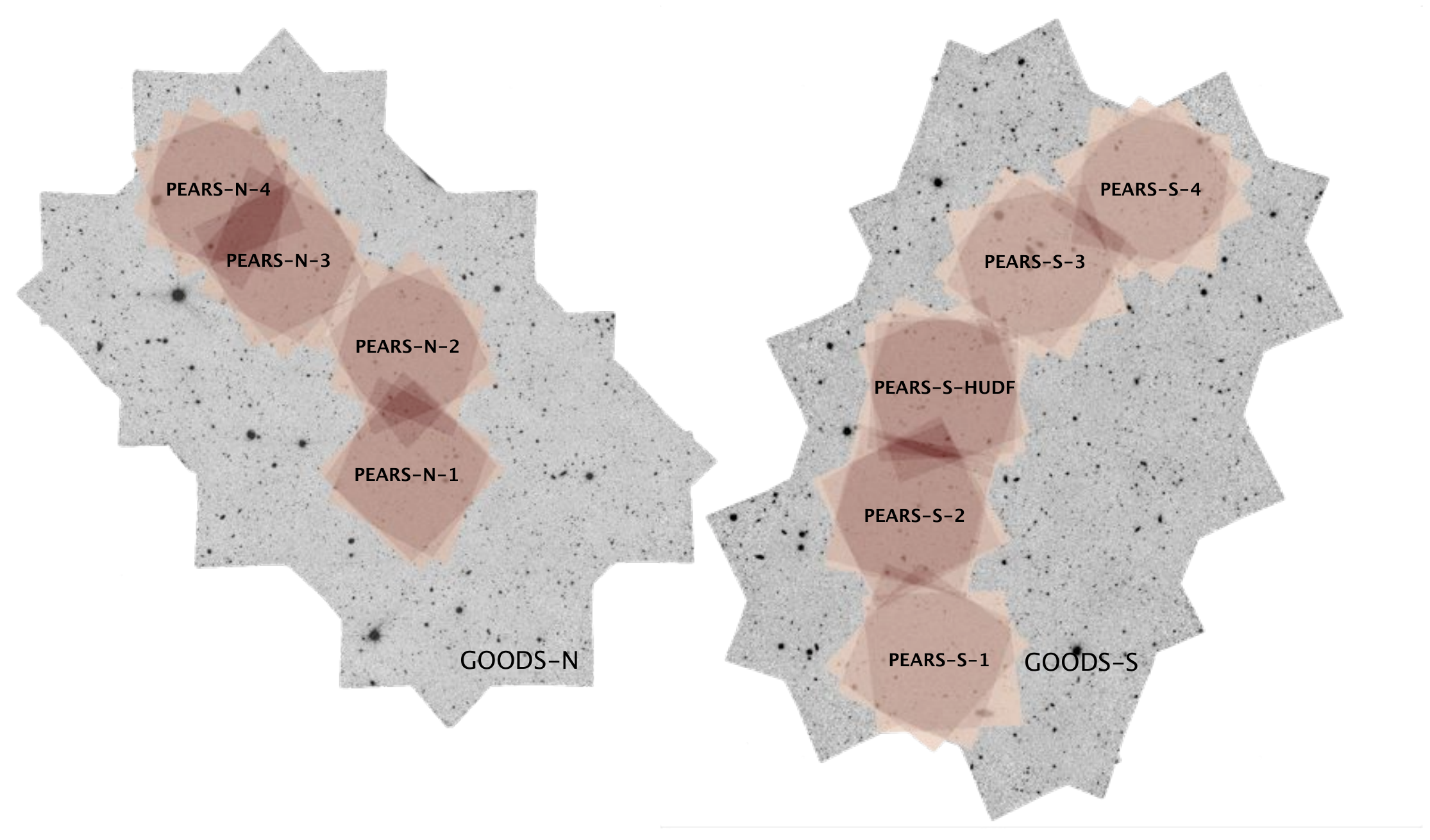}  
\caption{The location of the four PEARS-N (left) and five PEARS-S pointings (right) within 
the GOODS-N and GOODS-S fields. The fields are oriented so that North is up and East is 
to the left.  Each of the shown PEARS fields is approximately 200" arc second wide. Note 
that the total area where PEARS fields overlap is somewhat higher in  PEARS-N than in PEARS-S.
\label{PEARSNS}}
\end{figure}

\indent The PEARS observations were obtained as part of a large {\it Hubble Space
Telescope} ({\it HST}) proposal (200 orbits, Proposal 10530; P.I.: Malhotra).
The program used the Advanced Camera for Surveys (ACS) Wide Field Camera (WFC) 
in conjunction with its G800L grism filter.  The G800L has a resolution of 
{\it R} $\simeq$ 69-131, and provides wavelength coverage of 0.55-1.05 $\micron$
across the entire ACS/WFC field-of-view.  A total of nine fields 
(${\rm \approx 11.65\ arcmin^2}$ for each field) were 
observed for $\approx$40,000 s (20 orbits) each, split evenly between observations taken 
at different position angles (PA) on the sky (typically 3 per pointing). Multiple 
PA's are important for identifying and masking contamination from other sources
in the field, and for removing spurious pixels (e.g. cosmic rays, bad pixels, etc).
Four PEARS fields are within the GOODS-N field \citep{Giavalisco2004}. Five PEARS fields 
are within the GOODS-S field, with one PEARS field re-observing the GRAPES/HUDF field 
\citep{pirzkal2004}. The combined areas of the PEARS-N and PEARS-S are 50.24 and 
68.84 ${\rm arcmin^2}$, respectively. The PEARS fields and their location within the GOODS 
fields are shown in Figure \ref{PEARSNS}. Table \ref{table1} lists the PEARS fields 
positions and total exposure times. 

\section{Data Reduction and Analysis}\label{analysis}

\subsection{Detection of Emission Lines}\label{PEARS2D}
\indent PEARS emission lines were initially detected directly from combined high signal-to-noise ratio ACS grism slitless 
spectroscopic images. The method does not rely on, and is independent of, any imaging data or object catalog.
 The basic method used to identify emission lines in the 
PEARS data was described in \citet{straughn2008,straughn2009}. However, 
at that time, only the PEARS-S data were used.   We now present the 
full PEARS dataset (comprising of  PEARS-N and PEARS-S), which covers an area on the sky that is approximately twice as large.  
We have also employed a new, 
slightly refined version of our line identification pipeline, and are able to reach down to a slightly lower flux level than before. 
The detection algorithm, dubbed
``PEARS-2D,''  is very briefly summarized below and was applied to each of the individual 9 PEARS fields listed in Table \ref{table1}:\\
1)  All grism exposures obtained at the same PA on the sky are combined using the PYRAF task
{\tt MULTIDRIZZLE} \citep{Koekemoer2002}. This produces a high signal-to-noise ratio
image that is free of cosmic-ray and detector artifacts. This image is then 
smoothed using a $13\times3$\ median to produce a smooth high signal-to-noise ratio grism image of the field.  \\
2)  The smoothed image is  subtracted from the combined Multidrizzle image of the field. This step essentially removes 
the underlying continuum emission from the dispersed spectra.  The resulting continuum subtracted image is 
what we use to detect emission lines.  \\
3)  SExtractor \citep{sex} is used to  identify point sources, e.g. emission lines,
noise spikes, and detector artifacts  in the continuum subtracted image. This step generates a list of emission line candidates for each PA.\\
4)  Steps 1 though 3 are repeated separately for each available PA resulting in 3 or 4 (i.e. the number of available PAs) emission line candidates lists. A minimum 
of 2 PAs is required for the PEARS-2D method but additional PAs result in fewer false positives.\\
5)  Using a detailed knowledge of the instrument distortions and of the dispersion 
relation for the G800L ACS/WFC grism, and combining the data obtained using multiple PA's, 
we determine the location on the sky where each individual emission line originates (essentially where the different spectral traces cross when projected onto the sky), as well as determine the 
observed wavelength of each emission line. We derive a unique wavelength for each line in each of the available PA \citep[See Figure 2 of ][]{straughn2008}.\\

The PEARS-2D method was applied separately to each PEARS field listed in Table \ref{table1}, and a separate list of ELR candidates was
therefore generated for each of the nine PEARS fields, ignoring for now the fact that some of these fields overlap slightly.
The number of ELR candidates that was generated was controlled mainly by the detection threshold used to detect emission lines
with SExtractor, and by how much tolerance we allowed in the inferred ELR positions and therefore the observed wavelengths of the emission lines.
We adopted a detection of  $1.1\sigma$ per PA, which for a line detected independently using 3 different PA's corresponds 
approximately to a $2\sigma$ detection. We also allowed for a 3 pixel tolerance in the physical location of an ELR (accounting for 
an imperfect knowledge of the trace for the ACS grism), and a 100\AA\ tolerance in the wavelength that we derive for an emission line (again
accounting for an imperfect knowledge of the wavelength solution of the ACS grism). While the use of multiple PA's was quite effective at filtering 
spurious detections, our tolerance limits and aggressive detection threshold resulted in the detection of fainter emission lines, at the cost of some false-positive detections.\\

\indent Using the PEARS-2D method allowed us to  generate a list of ELR candidates for each PEARS 
field that did not rely on any pre-generated object catalogs or pre-selection of 
target galaxies.  We stress that a candidate ELR did {\it not} require the detection of  
a host galaxy in the field (The fraction of isolated ELR is discussed in Section \ref{lineid}).
 PEARS-2D has three immediate 
advantages over other methods that rely on observations taken at a single PA:
1) detecting extremely large EWs that would not be identified through 
more traditional photometric techniques;  2) deriving accurate locations of  
ELRs without assuming that the source is at the center of the host galaxy, and we can therefore
 identify multiple ELRs within a single galaxy; and 3) the PEARS-2D line wavelength calibration 
 is significantly more accurate.  Normally, the wavelength
reference point is tied to the location of the host galaxy (determined using a 
direct image taken in conjunction with the grism observations).  However, since ELRs can
be several half-light radii away (amounting to a non-trivial number of pixels). the
wavelength solution of the ELR is affected by this distance from the center of the host
galaxy. Every error of one pixel in the assumed position of the emission line feature 
results in a 40\AA\ systematic error in wavelength calibration, and therefore a significant redshift error.  For large galaxies with
multiple ELRs this can lead to errors on the order of several hundred \AA.  Using the PEARS-2D,
this error is avoided because the wavelength solution of the ELRs is determined 
{\it independent} of any information about the host galaxy.

\subsection{Extraction and Verification of Spectra}\label{extract}
\indent We extracted spectra of the ELRs candidates identified in Section \ref{PEARS2D}
using the  PEARS grism data reduction pipeline \citep[e.g.][]{pirzkal2009}. However, our pipeline was
modified, so that we used
 the positions of the ELRs candidates (i.e. the source of the emission features detected in our grism exposures), rather than a catalog of objects derived from the GOODS broad-band images.
The extraction and calibration of the spectra were performed 
with aXe package \citep{pirzkal2001,kummel2009}, using optimally weighted spectral extractions and an 
extraction width of 3$\times$ the measured emission region sizes.  Individual spectra were extracted
from single ACS grism exposures, and spectra taken at the same PA were then combined using aXeDrizzle \citep{kummel2009}. We thus
obtained multiple independent spectra for each of our ELR candidate (one per available PA).
Wavelength and flux calibration were performed by aXe using the STScI-provided calibration files for the ACS G800L grism.
Emission lines fluxes were measured using Gaussian fitting in each extracted spectrum, resulting in 3 to 4 independent measurements. We then 
computed the average line flux and its associated standard deviation value for each of our emission line candidates.
These line fluxes were corrected for Galactic reddening assuming 
values from \citet{cardelli1989}.  The corrections were negligible with (E(B-V)=0.012 mag for PEARS-N field and E(B-V)=0.0078 mag for PEARS-S field. \\ 
\indent  Our list of ELR candidates reached down to very low flux levels, and contained many false positives. We therefore had to vet
each ELR and its associated emission line candidates.
While we attempted to apply automatic techniques to accept or reject ELR and emission line candidates, we found it
useful to manually  vet  all spectra by eye.  Authors of this  paper manually examined 
and graded emission lines on a scale ranging from 0 (very poor) to 5 (very high).  Each emission line was graded a minimum of  3 times, and by at least
two different persons, and the average of this grade was then adopted.  A final grade of 2 was found to correspond to a 
marginal ${\rm \sim 2\sigma}$ detection of spectra obtained in at least two separate 
PA's. \\
We originally identified a total of 3705 emission-line candidates using the PEARS-2D method described in Section \ref{PEARS2D}.  
The visual quality assessment yielded a  sample of 1162 marginally detected emission lines (529 in PEARS-N and 633 in PEARS-S).  
As multiple emission lines were sometimes detected in an ELR, the final number of ELRs was 985 (451 in PEARS-N and 535 in 
PEARS-S).

\subsection{Emission Line \& Host Galaxy Identification}\label{lineid}
\indent Determining the nature of emission lines in the extracted spectra of a given ELR fell into two distinct categories: either multiple emission lines were detected, or only a single emission line was detected. \\
\indent The spectral dispersion of the G800L grism is $\sim 40$\AA\ pixel$^{-1}$ and is too low to 
resolve and identify close emission line pairs (e.g. \OIII\ and \Hb). However, in the redshift range of the PEARS survey  (0 $<$ z $<$ 1.5), 
there are pairs of widely separated emission lines that allow for both line identification (via the ratio of the observed wavelengths of the two lines) and 
redshifts to be determined.  The line pairs that we considered  were: \OII\ and \OIII; \OIII\ and \Ha; or \CIII\ and \CIV.  \\
\indent Most of the time spectra contained only one prominent emission line, so we had to rely on a comparison with 
photometric redshifts of the host galaxy from \citet{dahlen2010,dahlen2012}.  
While we noted above that the PEARS ELRs were selected independently,
we subsequently matched them with a host galaxy in the GOODS fields.  
We used the  public
ACS GOODS 2.0 data  to generate mosaics of the GOODS fields, and used SExtractor to 
 generate segmentation maps and object catalogs of galaxies for these fields. In the great majority 
of cases, the RA and Dec of a PEARS ELR clearly fell within the segmentation map of a 
galaxy.  In such cases the corresponding galaxy was assumed to be the host of the 
ELR, and it was assumed that the redshift of the observed emission lines to be the known photometric redshift for the host galaxy.
 Some ELRs ($\approx 6\%$ of objects with a significant emission line with a grade of 2.5 or larger) were found to lie beyond any galaxy segmentation maps.  In such cases, the
 photometric redshift of the GOODS object with the closest segmentation map was used (which is not necessarily the closest object in cases of large extended galaxies).   When comparing photometric redshifts to PEARS spectroscopic redshifts obtained using one of the pair of emission lines listed above, in 94\% of the cases, the redshifts were within the 95\% confidence regions given for the photometric redshifts of  \citet{dahlen2010,dahlen2012}.
When only a single emission line was detected in a spectrum, it was assumed to be either \Lya, \CIV, \CIII, \MgII, \OII, \NeIII, \OIII, \Hg, or \Ha, observed at the assumed GOODS photometric redshift value.  We identified the type of the emission line, and computed a spectroscopic redshift for that object simply by selecting the line type that produced the closest match to the observed wavelength.

It should be noted that some of the emission lines discussed in this paper are blended lines, but this should have little effect on our analysis 
,since weaker lines only  weakly bias the fluxes and redshifts that we derive. This 
is the case, for example, for \OIII\ which consists of two unresolved lines at $4959$\AA\ 
and $5007$\AA, \Ha\ at $6562$\AA\ which is blended with weaker  [NII] at 
$6583$\AA, while by \NeIII\ we actually refers to the stronger component at $3868$\AA.
Furthermore, some for the PEARS ELGs were found to contain more than one ELRs. Since these ELRs and the emission lines their spectra contained were
analyzed separately, this provided a way to check the consistency of the PEARS spectroscopic redshifts. In these cases, the redshifts  agreed to within z $=$ $\pm$ 0.01. \\
As noted earlier, some of the PEARS fields overlap slightly, thus these ELRs were  observed and analyzed independently as part of different PEARS fields. Nineteen ELRs were observed twice (15 in PEARS-N and 4 in PEARS-S; the higher number of duplicate observations in PEARS-N is the result of the larger amount of overlapping  between individual PEARS-N fields, as shown in Figure \ref{PEARSNS}). Comparing the observed emission lines wavelength, the emission line fluxes, and finally the redshifts, the errors were: $<\delta\lambda>=18$\AA; $<{\delta f \over f}>=8$\%; and $<\delta z>=0.003$, again demonstrating excellent consistency.
\indent For the remainder of this paper, we adopt a minimum emission line grade of 2.5 or greater (corresponding to a line flux limit of ${\rm \sim 1 \times 10^{-17}\ erg/s/cm^2}$), when deriving properties of \Ha, \OIII\ and \OII\ emitters (174, 401, and 167 emission lines, respectively).

\subsection{Spectroscopic vs. Photometric Equivalent Widths for ELRs at Large Radii}
\begin{figure}
\includegraphics[width =5in]{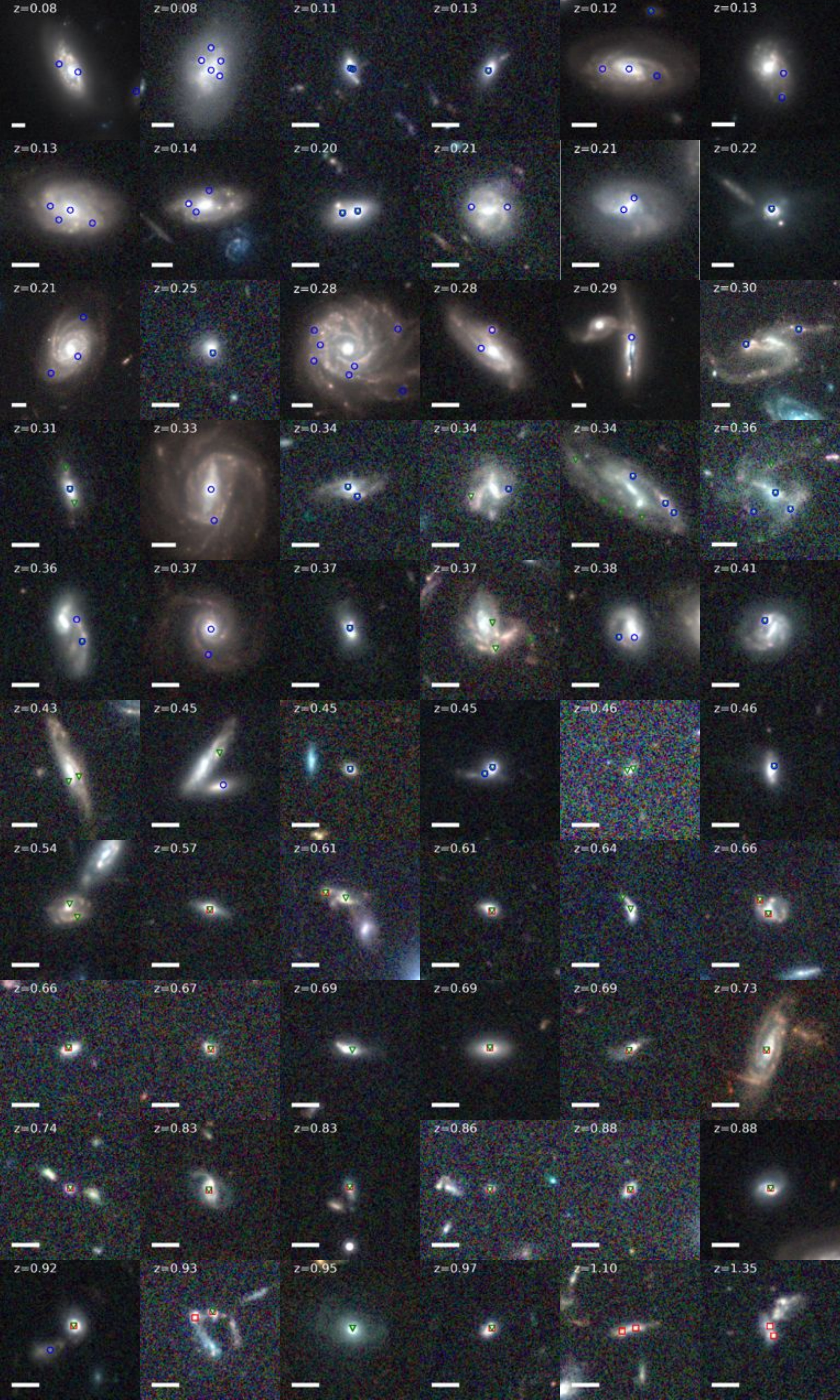}  
\caption{A sample of PEARS star-forming galaxies with their identified star-forming 
regions  by red squares (\OII), green triangles (\OIII), and blue circles (\Ha). The 
redshift is indicated at the top-left of each stamp image, and the one arc second scale is 
shown at the bottom-left of each stamp image. \label{stps}}
\end{figure}

\begin{figure} 
\includegraphics[width =7in]{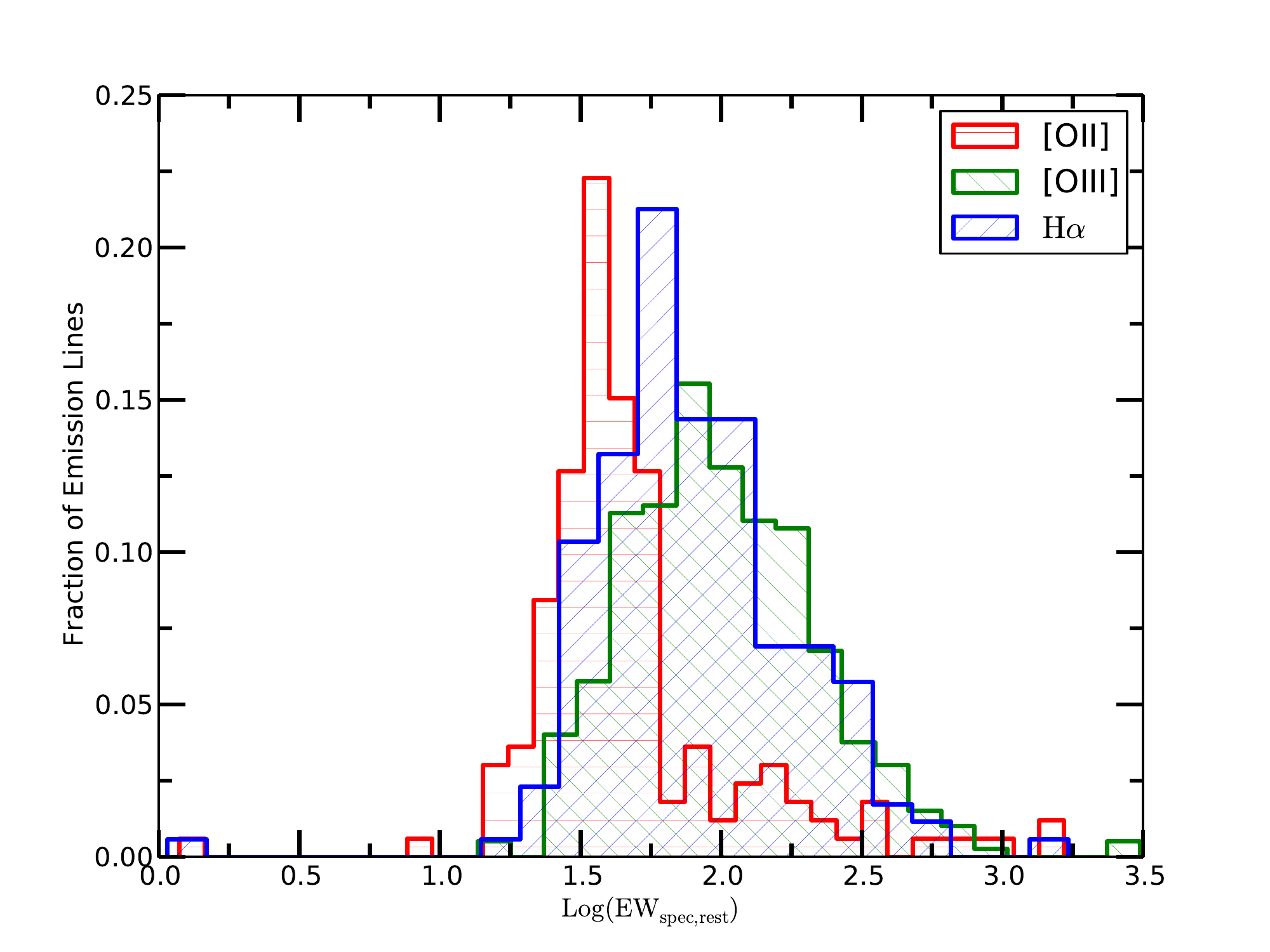} 
\caption{Rest-frame spectroscopic EWs of the PEARS-2D sample for emission-lines with a grade greater than 2.5. The median ${\rm EW_{spec}}$ 
we measured are 39\AA, 91\AA, and 71\AA\ for  \OII, \OIII, and \Ha\ emission lines, 
respectively. These lines act as proxies for the redshift ranges of $0 < z < 0.5$, $0.1 < z < 0.9$, and $0.5 < z < 1.5$, respectively.\label{EWsobs}}
\end{figure}

\indent One of the advantages of PEARS-2D is the ability to detect multiple ELRs within 
a single galaxy (see Figure \ref{stps}).   However, at progressively larger radii from 
the galaxy center, the contribution from the underlying continuum decreases.
Since spectra were extracted at these large radii using small extraction windows, 
the measured EWs are generally larger than what would be derived by simply comparing the 
measured line fluxes to the total underlying continuum of the host galaxy. The EWs derived
from narrow-band imaging surveys generally rely on the latter method.  To quantify any
potential differences, photometric EWs (${\rm EW_{phot}}$) were computed using the 
measured PEARS line flux and the measured total host galaxy broad band flux.  On average
the spectroscopic EW (${\rm EW_{spec}}$) was $\sim$ 3.5$\times$ (${\rm EW_{phot}}$). 
Histograms of the ${\rm EW_{spec}}$\ values for the \Ha, \OIII, and \OII\ are plotted in 
Figure \ref{EWsobs}.  We note that for the purposes of this paper an emission line
is reported as a positive EW.

\subsection{Blended Emission \OIII\ and \Hb\ Lines}\label{blend}
\indent The ACS G800L grism cannot separate the \OIII\ doublet (4959\AA, 5007\AA) and
\Hb\ (4681\AA).  In this paper, the  4959\AA, 5007\AA\ doublet is referred simply to \OIII. These three lines appear blended in the PEARS spectra, and \Hb\ can appear as an asymmetric feature
to the stronger \OIII\ line. To correct the \OIII\ fluxes for this effect, each of the lines were fit using separate components.  We assumed identical full 
width at half maximum (FWHM) and assumed a fixed wavelength separation for all three lines.
Based on this, we obtained estimates of the \OIII\ to \Hb\ lines ratio for the ELRs. 
: $ {f(H\beta) \over f(\OIII)} \approx 0.23\pm0.25$, which is consistent with 
the relative fluxes expected in star-forming galaxies \citep[e.g.,][]{Juneau2011}.

\subsection{Comparison with Previous PEARS-S Catalog}
The PEARS-S data were analyzed and published by \citet{straughn2009}, and are included as 
part of our analysis of the complete PEARS survey. While our improved analysis reaches down to 
fainter observed flux-levels, which results in a larger number of ELRs being detected, our emission line list contains in excess of 90\% of the emission lines listed in   \citet{straughn2009}. However, the  number of \Ha, \OIII, and \OII\ emission lines (with a grade of 2.5 or above) are 1.2, 1.6, and 2.7 times higher, respectively, than in the \citet{straughn2009} catalog. This is a direct result of  our ability to reliably detect fainter emission lines in these data, which  particularly affects the intrinsically fainter \OIII\ and \OII\ emission lines. The use of optimal weighted extraction, as opposed to narrow box extraction  \citep[increasing signal-to-noise ratio at the expense of flux completeness, as was the case in][]{straughn2009}, results in more reliable flux measurements -- without a need for an aperture correction -- of these emission lines and the lines fluxes  measured are on average twice as strong as those listed in \citet{straughn2009}.

\subsection{Completeness Simulations}\label{completeness}
\begin{figure} 
\includegraphics[width =7in]{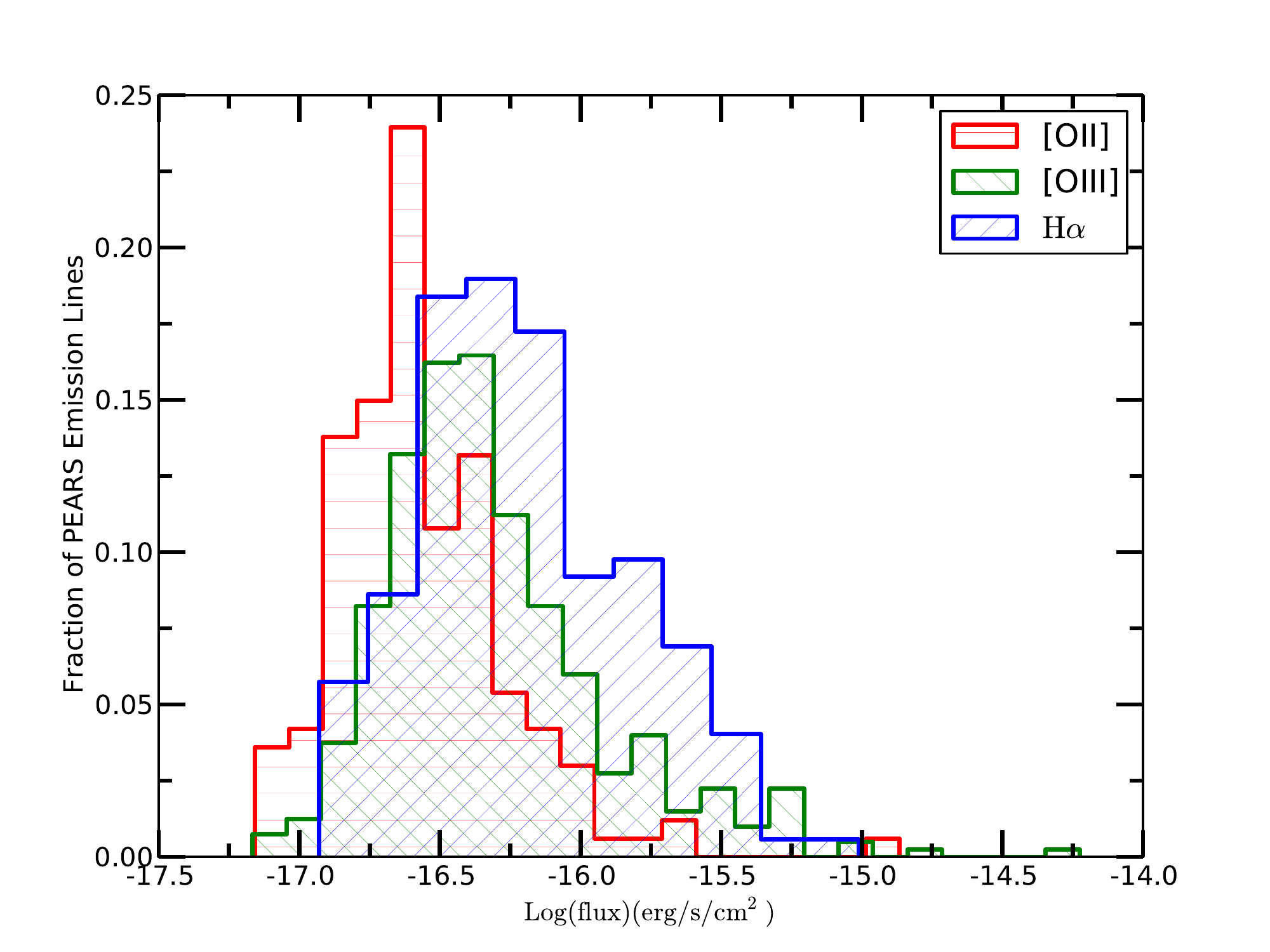}  
\caption{Distributions of observed line fluxes in the PEARS-2D sample for emission-lines with a grade greater than 2.5, uncorrected for 
completeness or dust extinction. These lines act as proxies for the redshift ranges of $0 < z < 0.5$, $0.1 < z < 0.9$, and $0.5 < z < 1.5$, respectively.
\label{allfluxes}}
\end{figure}

\begin{figure} 
\includegraphics[width =7in]{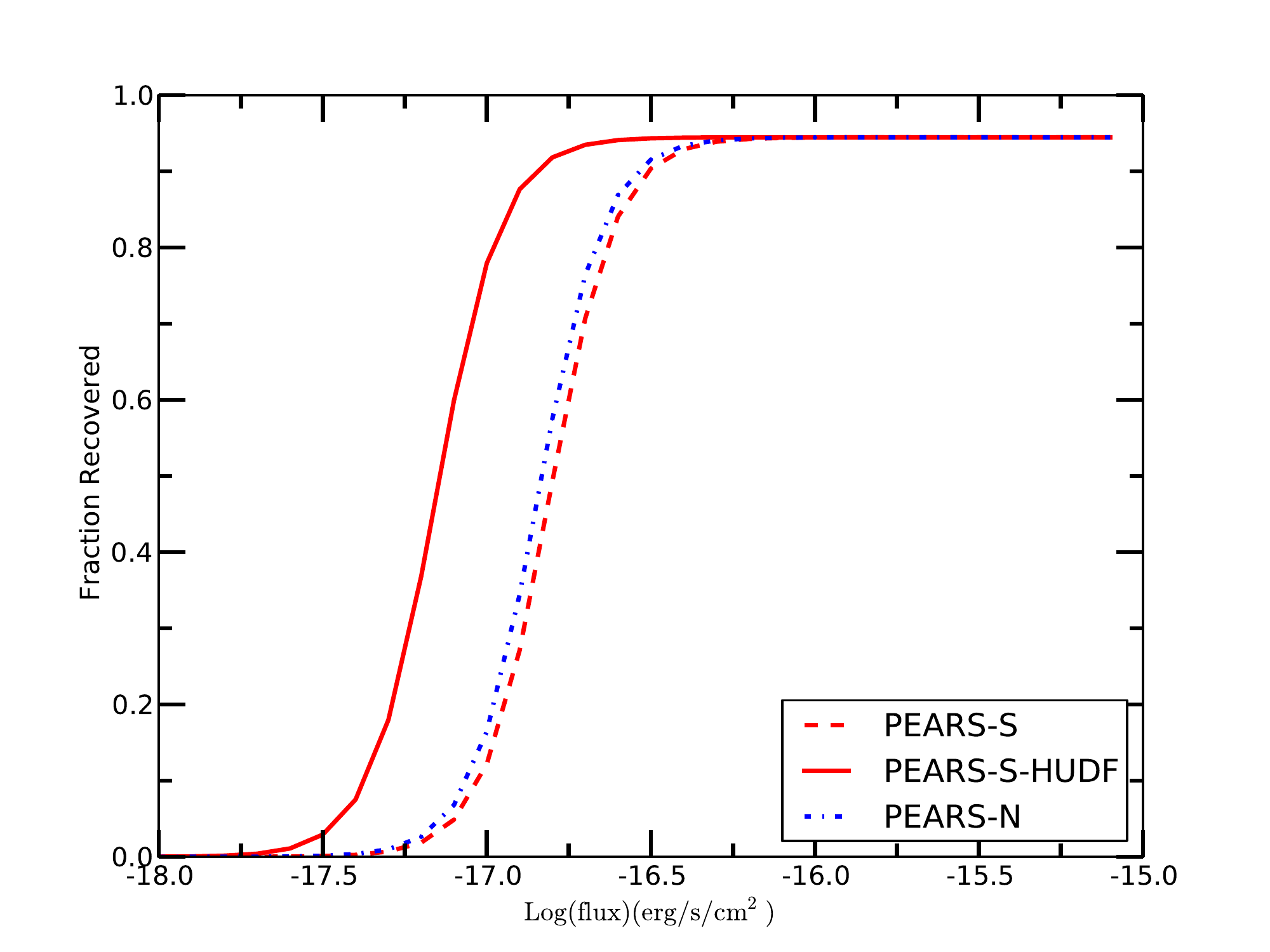} 
\caption{PEARS sensitivity to emission-line flux. Based on our simulations, we can reliably ($>85\%$) detect emission lines with fluxes greater than 
$10^{-16.5}$\ (${\rm 3 \times 10^{-17} erg/s/cm^2}$) over the whole PEARS field while the PEARS-S-HUDF 
field, which was observed twice as long as each of the other 8 PEARS fields, reaches line 
fluxes approximately 1.4 times fainter. The $50\%$ line-flux completeness limit is approximately $10^{-16.7}$\ (${\rm 2 \times 10^{-17} erg/s/cm^2}$).
\label{correction}}
\end{figure}

Table \ref{linefluxes} lists the median and average line fluxes for \Ha, \OIII, and \OII\ 
that have a strong detection (PEARS grade of at least 2.5). 
 In Figure \ref{allfluxes}, we 
show the distributions of the line fluxes for these three lines. The histograms
are plotted as a fraction of the total for each line.  Figure \ref{allfluxes} demonstrates
that the PEARS-2D line fluxes peak at values of $\sim$ ${\rm  10^{-17}\ erg/s/cm^2}$. 
The ACS G800L grism has an approximately flat sensitivity from 
${\rm \approx6000\AA\ to\ 9500\AA}$, but our ability to recover emission 
lines from the two dimensional dispersed images needs to be carefully evaluated before 
we can say anything about our completeness limits and the volume density of these sources. This is particularly important
because it must be verified that our sample is not biased by the host galaxy sizes, which vary as a function 
of redshift (i.e. all lines types are detected equally well).
We determined the 
PEARS-2D detection limits using extensive end-to-end Monte Carlo simulations.  These
steps are briefly outlined here:  First, we started with the real individual PEARS ACS/WFC G800L exposures and 
artificially added a random distribution of simulated ELRs (and simulated emission lines)
to these data.  We allowed for a wide ranges of line fluxes, host galaxies, and redshifts in these simulations, and   also allowed for random placement of multiple ELRs in each galaxy. Each of the nine PEARS field was simulated ten times, each time adding ELRs to 100 galaxies.\\
\indent  Next, the simulated data were processed and identified using exactly the same 
procedures used for the real observations.  The simulated spectra were then extracted using aXe, and line 
fluxes measured, as we described in Section \ref{extract}.  
Comparing input and output line lists, we found that the PEARS wavelength sensitivity 
is very similar to the inverse sensitivity of the ACS G800L grism with a sharp 
cutoff below 5500\AA\ and above 9500\AA. This sets the redshift ranges of the PEARS 
survey for the \Ha, \OIII\ and \OII\ lines to be $0<z<0.5$, $0.1<z<0.9$, $0.5<z<1.5$, 
respectively.  PEARS was mainly limited by the intrinsic emission line flux,
  ELRs  containing fluxes as low as 
${\rm 10^{-18}\ erg/s/cm^2}$ could be detected, and that we could detect more than 85\% of 
emission lines with flux greater than  ${\rm \sim3 \times 10^{-17}\ erg/s/cm^2}$ and with ${\rm EW > 50\ \AA}$. 
This is consistent with the observations of observed line fluxed shown in Figure \ref{allfluxes}.
The exact fractions of lines recovered as a function of observed line flux for the 
PEARS-N, PEARS-S and PEARS-S-HUDF field (which is twice as deep as the other PEARS-S fields) 
are shown in Figure  \ref{correction}.

\subsection{Methods for Computing Luminosity Functions and SED Fitting}
Here,  the methods adopted to determine the PEARS emission line luminosity functions for \Ha, \OIII\ and \OII\ are briefly described. We also outline how stellar masses were derived
for the ELGs containing the ELRs identified in the previous Sections. Finally, we discuss the different approaches adopted to account for internal dust corrections when computing the PEARS line luminosity functions. 
\subsubsection{The \vmax method}\label{vmax}
\indent This widely used method does not assume a shape for the luminosity function $\Phi(L)$. 
However, one disadvantage is that it requires the data to be binned.
The number of bins can impact the results. In this paper, the number of bins was 
determined using the Freedman-Diaconis rule \citep{freedman1981}, whereby the bin-size is selected
to be ${\rm 2\ IQR(x) n ^{-{1\over3}}}$, where IQR is the interquartile range of the data and $n$\ is the number of 
data points in the sample.\
Using the \vmax\ method, the luminosity function is computed using the following
formula:

\begin{equation}
\Phi({\rm \log L_i}) =  {1 \over{\Delta {\rm \log L}} } \sum_j {1\over{f(z_j,L_j) V_j}}
\end{equation}

\noindent where: $\left|{\log L - \log L_i}\right| < {1\over 2} \Delta \log L$;
$\Delta \log L$ is the bin-width; $V_j$ is the maximum volume within which
object $j$ (observed to have a line flux of $l_j$ and to be at the redshift of $z_j$) would be detected in our survey; and $f(z_j,L_j)$ is the incompleteness 
$f(l)$
(see Section \ref{completeness}), remapped into absolute luminosity space $L$ given the object's 
redshift $z_j$, and is defined as:

\begin{equation}
V_j = {\Omega \over {4\pi}} \int_{z_{j,{\rm min}}}^{z_{j,{\rm max}}} {R(z) dV_c(z) \over dz} dz
\end{equation}

\noindent where: $\Omega$ is the solid angle of our survey (sr); $V_c(z)$ is the 
cosmological comoving volume element at redshift $z$ (in ${\rm Mpc^3}$); and $R(z)$ is the normalized grism response function expressed as a function of object redshift. Given the redshift range $\left\{ z_{\rm l}, z_{\rm h}\right\}$ at which a given emission line can be observed by the ACS grism (i.e. observed at wavelengths ${\rm 6000\AA < \lambda < 9000\AA}$), the minimum redshift, $z_ {j{\rm ,min}}$ is $z_{\rm l}$ while the maximum redshift $z_{j,{\rm max}} = \min(z_{\rm h},z_{\rm faint})$, where $z_{\rm faint}$ is the maximum redshift at which a line with luminosity L would remain above our minimum line detection threshold $l_{\rm threshold}$. Hence, $z_{\rm faint}$ is the redshift corresponding to the distance of ${\rm D_L}(z_{ j})\sqrt{l_{j}/l_{\rm threshold}}$, where ${\rm D_L}(z_{\rm j})$ is the luminosity distance of object ${ j}$.

\subsubsection{STY method}\label{STY}
\indent The STY (Sandage, Tammann, Yahil) method \citep{Sandage1979} is another commonly used method for estimating the luminosity function.  In this one,
one assumes that $\Phi(L)$ has the form of a Schechter function \citep{Schechter1976}:
 
\begin{equation}
\Phi(L) dL= \Phi_* ({L \over L_*})^{\alpha}  \exp{(-{L \over L_*})} {dL\over L_*} \label{phi}
\end{equation}
\noindent which is characterized by the three parameters $\alpha$, $\Phi_*$, and $L_*$. 
Following  \cite{Sandage1979}, the probability of observing a given object $j$ 
at redshift $z$ with a luminosity $L_j$,  is then:

\begin{equation}
p(L_j,z_j) = {\Phi(L) f(z_j,L_j) R(z_j) \over \int_{L_{faint}}^\infty \Phi(L) f(z_j,L)   dL } \label{p}
\end{equation}

\noindent The joint likelihood can then be computed for the whole group of observed lines:

\begin{equation}
L = \Pi_j p(L_j,z_j) \label{L}
\end{equation}

\noindent From this, can then determine values of $\alpha$ and $L_*$ that maximizes this 
likelihood. The overall normalization constant $\Phi_*$ cannot be determined this way, 
because it cancels out in Equation \ref{p}.  In this paper, we determined the values of 
$\alpha$ and $L_*$ by maximizing Equation \ref{L} using a Monte Carlo Markov Chain 
approach. This allowed us to determine the most likely values of these two parameters, 
as well as 95\% credible intervals for these parameters.  $\Phi_*$ was computed by 
integrating Equation \ref{phi} and normalizing the result, so that matched the number of
detected objects.  $\Phi_*$ was computed for each combination of $\alpha$ and $L_*$ in our 
Markov Chains to produce 95\% credible intervals for the parameter $\Phi_*$.

\subsubsection{Host galaxy SED fitting}\label{hostSED}
\indent Properties of the host galaxies of the PEARS ELRs were estimated by fitting model 
Spectral Energy Distributions (SEDs) to their broadband photometric colors.  The 
photometric values were taken from the TFIT GOODS measurements \citep{Papovich2002}, which include 10 and 12 
photometric bands in the GOODS-N and GOODS-S, respectively \citep{Laidler2007,Grogin2012}.  
The photometry spans the observed UV ({\it U}-band) through thermal-IR based on VLT, 
{\it HST} and {\it Spitzer} observations. The majority ($95\%$)  of the PEARS ELGs are detected in the Spitzer data and
these observations therefore probe the rest-frame optical increasing the robustness of our SED fitting. The actual SED fitting was done using our own
Monte Carlo Markov Chain SED fitting code (${\rm \pi MC^2}$) \citep{pirzkal2012} to 
obtain estimates of the stellar masses, extinction, and ages of the host galaxies.   
${\rm \pi MC^2}$ is a far more robust method of SED fitting than the standard $\chi$$^{2}$  
algorithm because it provides a proper treatment of both error propagation
, and a computation of confidence levels. A more detailed explanation of MCMC can be found in 
\citep{pirzkal2012} and references therein.

\begin{figure} 
\includegraphics[width =7in]{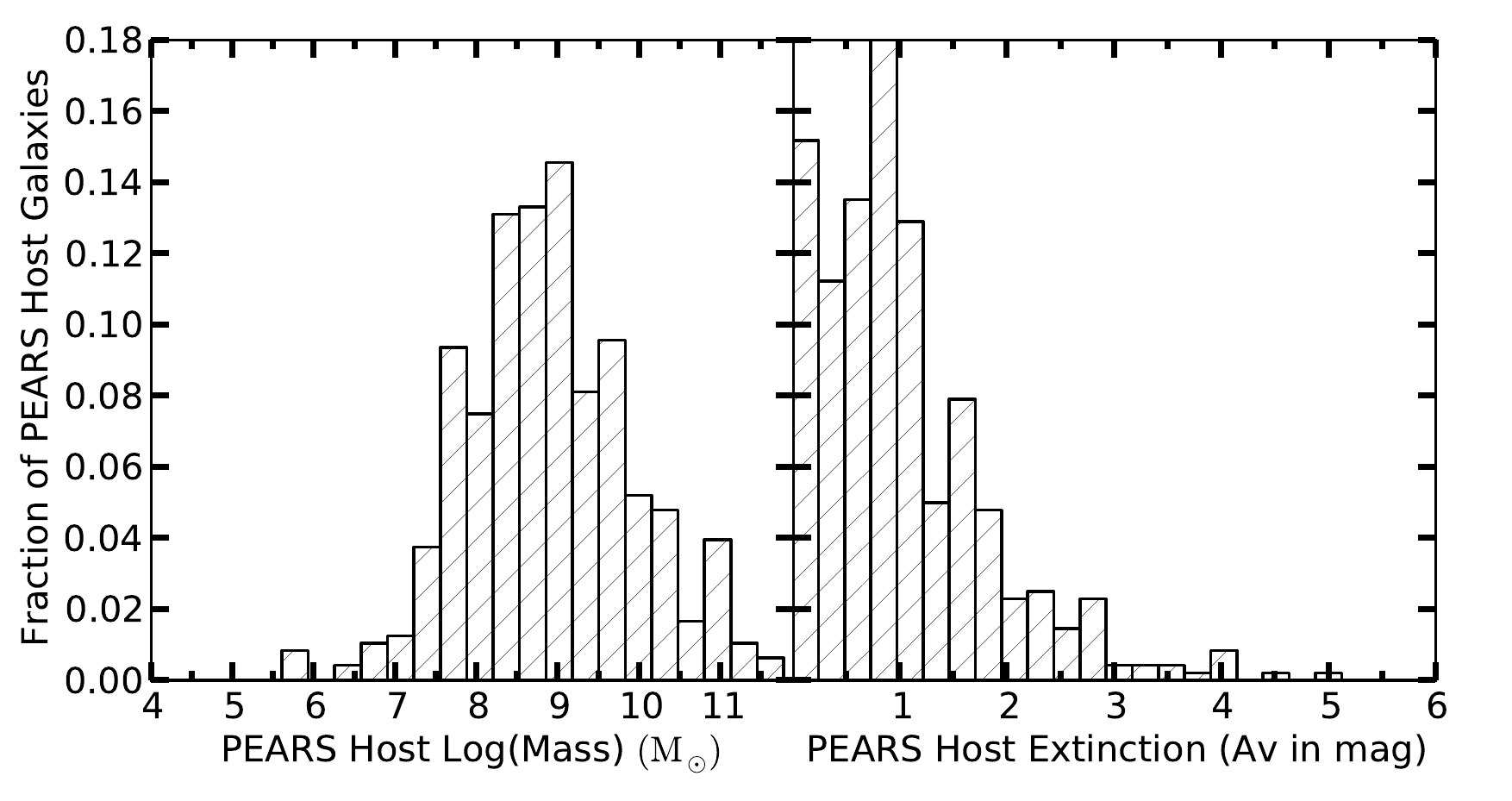}  
\caption{Left panel: Histogram of the PEARS emission-line galaxy stellar masses 
(in $M_{\sun}$) as determined from the SED fitting.  We derive a mean stellar mass of 
${\rm Log(mass) = 8.85 \pm 1.03\ M_\sun} $\ (1$\sigma$). Right panel: Histogram of the PEARS 
emission-line galaxy extinction (Av in mag) as determined from SED fitting. A mean extinction of ${\rm Av = 0.88 \pm 0.93}$\ mag (1$\sigma$) is derived. 
\label{masshist}}
\end{figure}

  A simple stellar population model (i.e. single burst) with BC03 \citep{bc03} templates and a Salpeter IMF were used. While the choice of IMF and 
input models \cite[e.g. BC03 or ][]{ma05} can affect derived
stellar masses, the effects are not the same at all redshifts.  The detailed 
simulations presented in \citet{pirzkal2012} show that for the redshift range of interest 
here, stellar mass estimates from different models are consistent with each other to
within a factor of a few. The typical uncertainty in our stellar mass estimates is $\sim 0.25$\ dex while
the typical uncertainty in our extinction estimates is $\sim  0.35$.
Other parameters obtained from SED fitting (e.g.
extinction, metallicity, and ages of the stellar population) are significantly
more uncertain.  For the purposes of this paper, we are primarily concerned with 
stellar mass, and to some extent extinction. In Figure \ref{masshist}, we show 
the distribution of stellar masses and extinctions for the host galaxies of the PEARS 
emission line sample. The mean stellar masses of the host ELGs are 
${\rm Log(mass) = 8.85 \pm 1.03\ M_\sun}$. We also estimate that the continuum extinction  
is relatively low, with an average value of ${\rm Av = 0.88 \pm 0.92}$\ mag, listing the $1\sigma$\ dispersion in each case.

\subsubsection{Internal Dust corrections to Luminosity Functions}\label{dustme}
\indent Using the current PEARS data, there was no way to directly infer the amount of
internal (to the host galaxy) dust attenuation affecting the line luminosities.
Therefore, three methods were tested for approximating  dust corrections to the luminosity functions, and compared to 
the ones from \citet{Ly2007}.  The first dust correction used attenuation 
values from the individual SED fits to the host galaxies.  The second method  relied on applying an average extinction value 
of  ${\rm A_{H_\alpha}=1.0\  mag}$  (corresponding to ${\rm A_{[OII]}=1.88\  mag}$ and ${\rm A_{[OIII]}=1.36\  mag}$), 
as is commonly done in SED fitting \citep[e.g.,][]{hopkins2004,takahashi2007}. The third method relied on a dust correction based on  the somewhat more sophisticated 
luminosity dependent dust extinction of  \citet{hopkins2001} 
\\
\indent While the two first approaches are straightforward, they are a rather coarse attempt at applying dust-corrections. Indeed, these approaches do not allow for the extinction values within ELRs to be 
different than the host galaxy, and  do not allow for the possibility that ELRs might contain a different amount of dust than the host galaxy. The typical nebular extinction is greater than the stellar continuum 
emission extinction. Applying SED derived extinction values to the PEARS emission lines are therefore likely to be too low \citep{calzetti2000}. The luminosity dependent approach empirically attempts to circumvent this limitation. In this case, the amount of dust correction is correlated
with the measured line luminosities. This method, unlike the first two, could in principle affect  the shape of the luminosity functions.
We implemented the luminosity correction following the equations listed in \citet{hopkins2001} and re-deriving equation 5 therein for the specific wavelength of
either \Ha, \OII, or \OIII.
\\
\indent However, the three methods of correcting for dust have only limited effect on the resulting luminosity functions. They simply shift the luminosity 
functions by a fixed amount without affecting the slope ($\alpha$) at all (as is the case when using the first two methods), or only alter the slope ($\alpha$)
 slightly (as is the case when applying a luminosity dependent dust-correction).  As expected, the \OII\ lines 
were more affected by dust than the \OIII\ and \Ha\ lines. However, in total, 
for \Ha, \OIII\ and \OII\ the slopes varied by only $\sim$ 0.1 on average, using either of the three methods discussed above. Hence, we conclude that the effect of dust correction on the slope of the luminosity functions is therefore negligible to within the statistical fitting error in $\alpha$.

\section{Results}
\subsection{The Emission Line Regions}\label{ELR}
\subsubsection{Star-Forming Galaxy Density}
\indent Using the uncorrected PEARS lines listed in Table \ref{linefluxes}, we can
compute a space density of star-forming galaxies (SFGs) --- as measured by the PEARS survey --- and compare it
to previous ACS grism based surveys. The star-forming galaxy density at $0.3 < z < 1.3$ is estimated to be 
${\rm4.5 \times 10^{-3}\ Mpc^{-3}}$.  This is in complete agreement with previous  ACS grism  pure parallel
surveys, such as the one described in \citet{drozdovsky2005}.  

\subsubsection{Luminosity Functions}\label{luminosity}
\begin{figure} 
\includegraphics[width =7in]{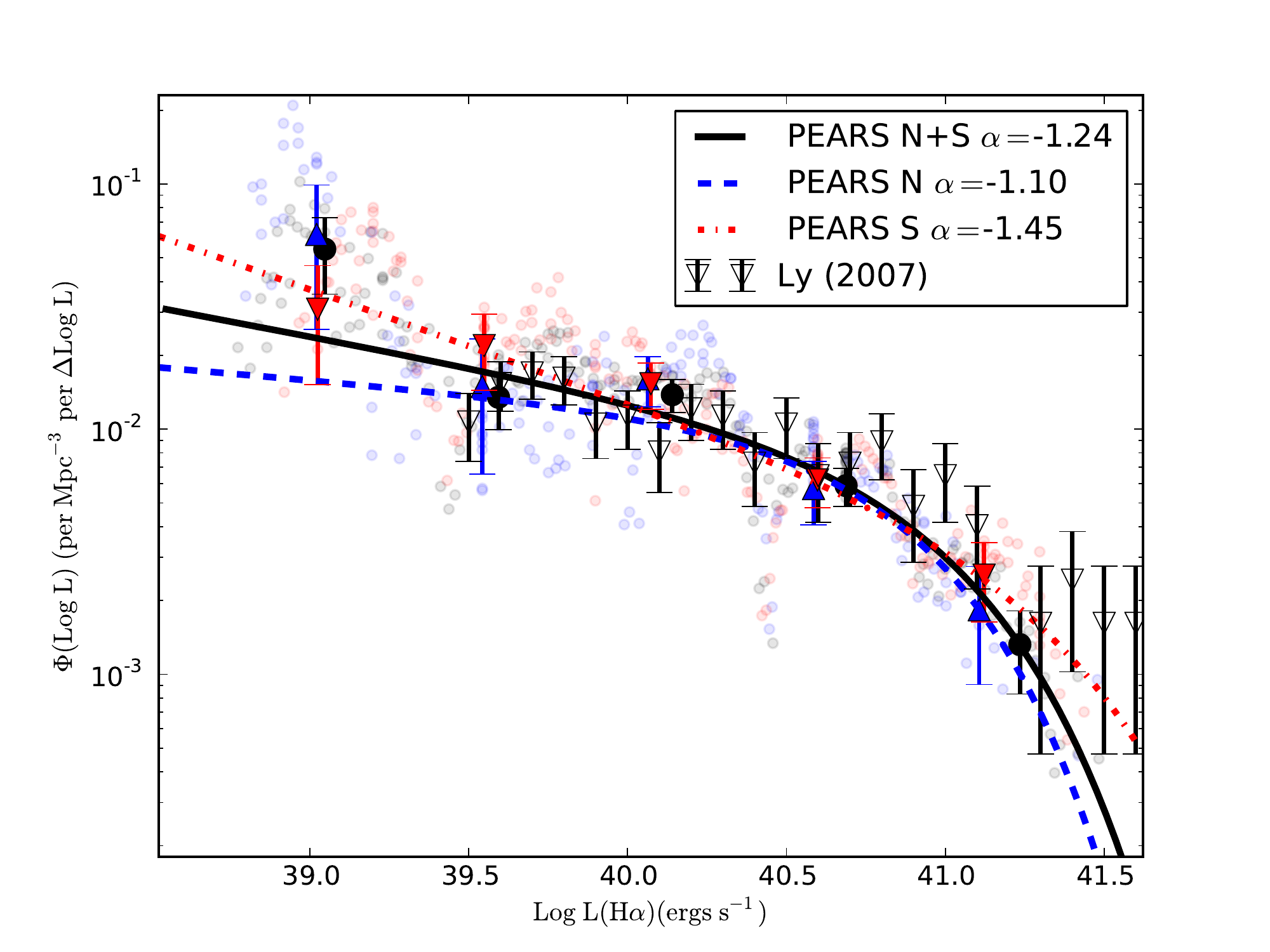} 
\caption{PEARS \Ha\ luminosity function  at $0<z<0.5$. We  show  the \vmax\ results for the full PEARS,  
PEARS-N and PEARS-S in black circles, blue upright triangles, and red downward triangles, 
respectively. The fits to the \vmax\  results are shown by the solid black line, blue 
dashed line, and red dash-dotted line, for  PEARS, PEARS-N and PEARS-S, respectively. No 
significant differences are found between the PEARS-N and PEARS-S fields. We also plot the 
sample of $z=0.4$\ \Ha\ emitters from \citet{Ly2007}, also with no dust correction and excluding 
objects with ${\rm EW<50\AA}$ from their sample, to better compare results, 
and illustrate how PEARS reaches to fainter luminosities for objects that were selected in similar ways.
\label{Halumfctdustns}}
\end{figure}

\begin{figure} 
\includegraphics[width =7in]{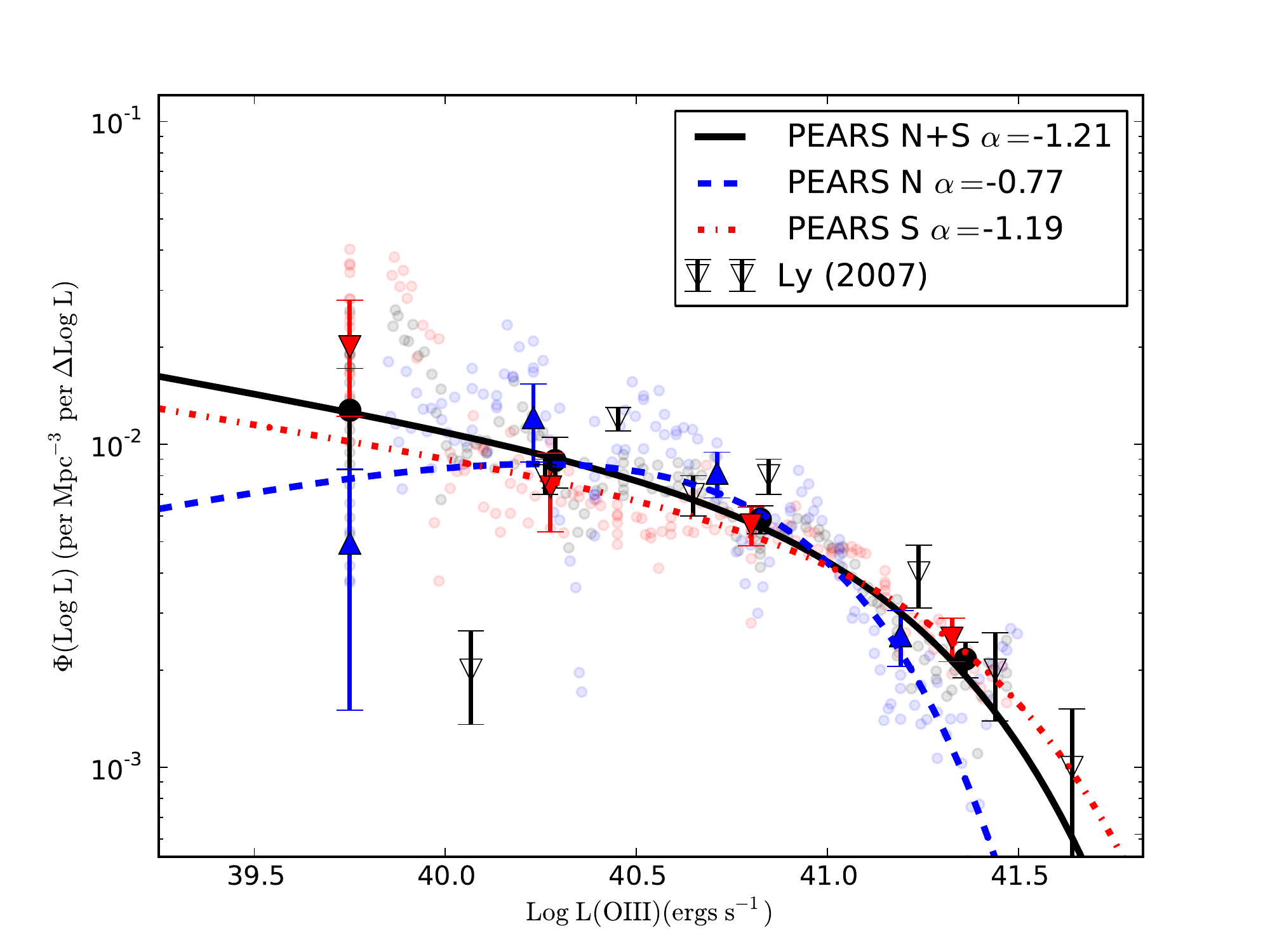} 
\caption{PEARS \OIII\ luminosity function  at $0.1<z<0.9$. We  show  the \vmax\ results for the full 
PEARS, PEARS-N and PEARS-S in black circles, blue upright triangles, and red downward 
triangles, respectively. The fits to the \vmax\  results are shown by the solid black 
line, blue dashed line, and red dash-dotted line, for  PEARS, PEARS-N and PEARS-S, respectively. No
 significant differences between the PEARS-N and PEARS-S fields are detected. We also plot the 
sample of $z=0.6$\  \OIII\ emitters from \citet{Ly2007}, also with no dust correction and excluding 
objects with ${\rm EW<50\AA}$ from their sample to better compare our results 
and illustrates how PEARS reaches to fainter luminosities for objects selected in similar ways.
\label{OIIIlumfctdustns}}
\end{figure}

\begin{figure} 
\includegraphics[width =7in]{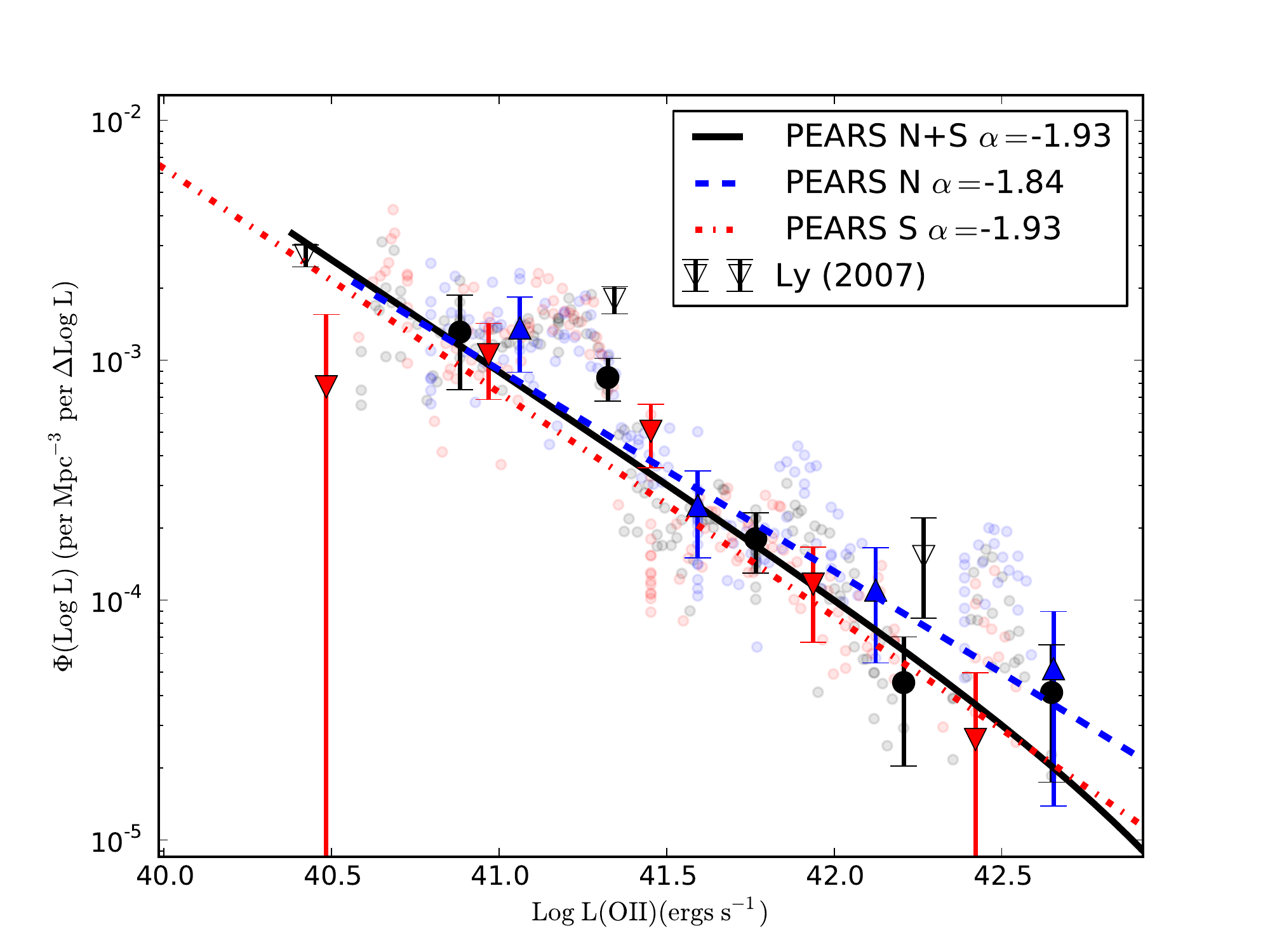}  
\caption{PEARS \OII\ luminosity function at $0.9<z<1.5$. We  show  the \vmax\ results for the full PEARS,  
PEARS-N and PEARS-S in black circles, blue upright triangles, and red downward triangles, 
respectively. The fits to the \vmax\  results are shown by the solid black line, blue 
dashed line, and red dash-dotted line, for  PEARS, PEARS-N and PEARS-S, respectively. No 
significant differencesare found between the PEARS-N and PEARS-S fields. We also plot the sample 
of $z=0.9$\ \OII\ emitters from \citet{Ly2007}, also with no dust correction and excluding objects 
with ${\rm EW<50\AA}$ from their sample, to better compare  results. 
\label{OIIlumfctdustns}}
\end{figure}

\indent We computed the luminosity functions for our \Ha, \OIII, and \OII\ samples, using 
both the \vmax method and the STY methods described in Sections 
\ref{vmax} and \ref{STY}.  The luminosity functions presented are for ELGs and not simply ELRs.  We computed
an integrated line luminosity for each galaxy by summing up the contributions from all ELRs in each galaxy. 
The non dust-corrected luminosity functions, computed using 
the \vmax\ method, are shown in Figures \ref{Halumfctdustns} to \ref{OIIlumfctdustns}. 
In these figures, the new measurements are compared to those of \citet{Ly2007}, shown with 
open triangles and also uncorrected for dust.  One additional constraint
to the data from  \cite{Ly2007} was added, namely that the comparisons were made only with objects with EWs $>$ 50\AA.\\
\indent The results from PEARS agree fairly well with earlier results, although
PEARS probes lower line luminosities for \OIII\ and \Ha.  The \vmax\  results from PEARS-N
and PEARS-S are plotted separately in Figures  \ref{Halumfctdustns} to \ref{OIIlumfctdustns}.
   The differences between these two large and independent fields 
are well within the statistical errors. Table \ref{lumfcttable} summarizes the results from fitting
the luminosity functions to each of the emission lines in each field separately, as well as 
PEARS-N and -S together using both the \vmax\ and STY methods.
  Table \ref{lumfcttable} also includes the associated 95\% credible intervals. \\
\indent When using the \vmax\ method, as  noted in 
Section \ref{vmax}, the choice of bin size is important. We illustrate the effect of various bin 
sizes by showing (using light shaded circles) the luminosity functions that we compute while allowing the bin
sizes to vary. As one can see, the effect of bin size has an immediate effect on the values that we derive at
a given luminosity. The luminosity derived with the optimal bin sizes is shown using solid symbols. The error bars associated with individual points
were derived using a few thousands bootstrapping iterations.
\\
\indent Here, we note that the STY method produces slightly different
luminosity function slopes than \vmax\ for \OII, although the two methods are consistent
with each-other.  The differences between the two are likely due to the 
limited number of sources over a wide redshift range.  This underlines the 
difficulties in obtaining luminosity function estimates from a limited numbers of sources,
as well as the impact of using different methods..  The results shown in Figures \ref{Halumfctdustns} to \ref{OIIlumfctdustns} 
are also in general consistent with those of \citet{Ly2007}.  
The fraction of AGN in our sample is low. Comparing our catalog to x-ray detected sources in the
GOODS areas \citep{alexander2003}, we estimate that AGN contamination is  $\sim 3\%$.

\subsubsection{The Spatial Distribution of ELRs}
\begin{figure}
\includegraphics[width =7in]{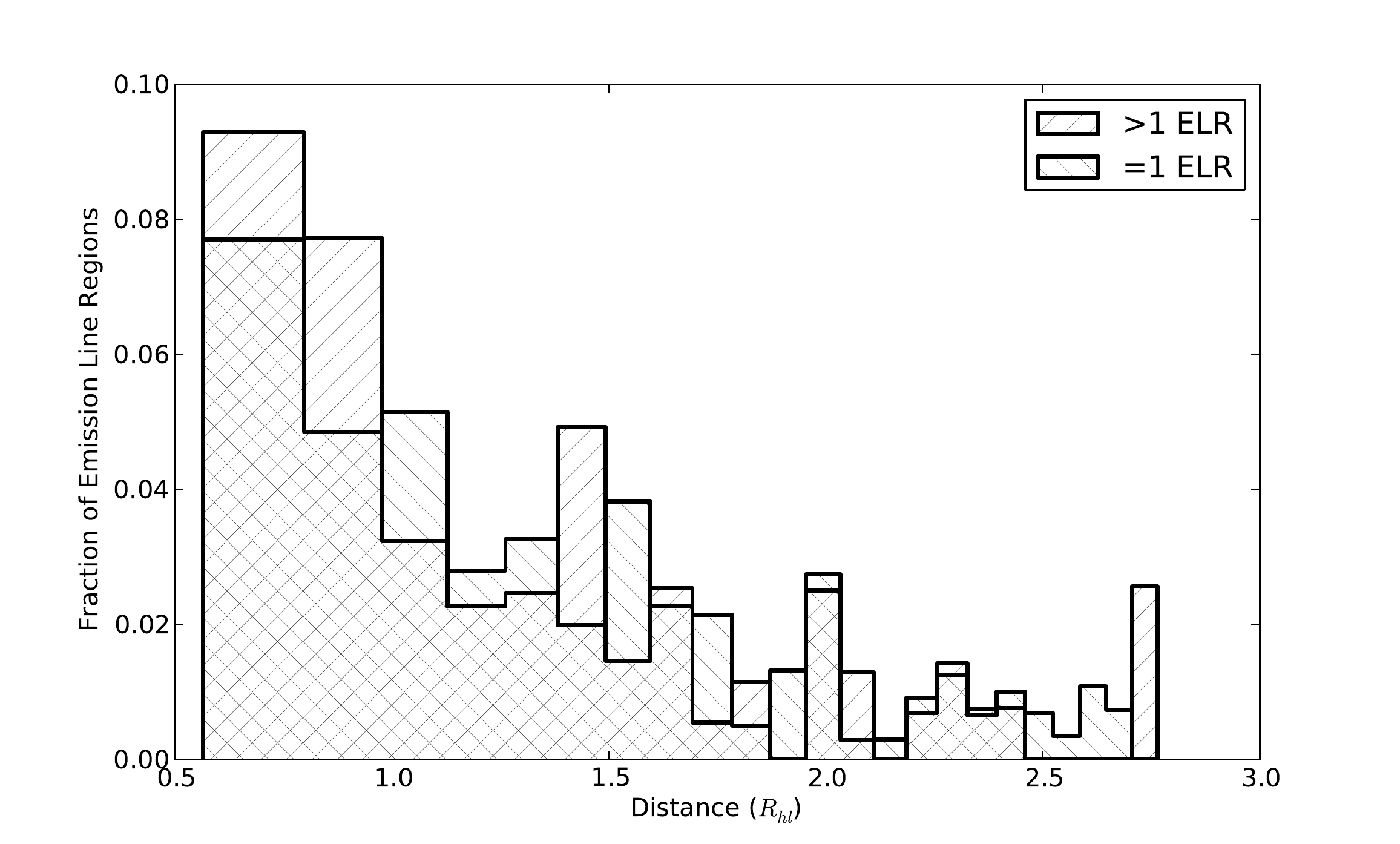} 
\caption{Histograms of the distance of the PEARS  ELRs from the center of their host 
galaxies, measured in units of $R_{hl}$. Bins sizes were selected to correspond to equal areas. We show the distribution of 
ELRs in ELGs where only one ELR was identified as well as the distribution of ELRs in 
ELGs, where more than one ELR was identified. A K-S test p-value of 0.49 indicates that no significant differences exist between the two distributions.
\label{ReHist2}}
\end{figure}

\indent A major difference between PEARS-2D and other ELG studies is that we are
able to detect the presence of multiple ELRs within a single galaxy. A breakdown
of the sample shows that 69\% of the ELGs contain a single ELR; 24\% contain two 
ELRs; 4\%  contain three ELRs; and 3\%  contain four or five ELRs.  
Comparing the location of these ELRs, as parametrized by their distance to the center of the host ELG normalized by the half-light radius ($R_{hl}$) of the host ELG, allows us to compare the distribution of single emission-line regions to the distribution of multiple emission-line regions.  Figure \ref{ReHist2} shows that there is no indication of any strong differences between the two samples. In both cases, the emission-line regions appear to be predominantly located around one half-light radius away from the center of the ELG. A Kolmogorov-Smirnof test (KS) p-value of 0.49 was computed, which
fails to demonstrate that these distributions are statistically different.

\subsubsection{Star Formation Rates of ELGs}
\begin{figure}
\includegraphics[width =7in]{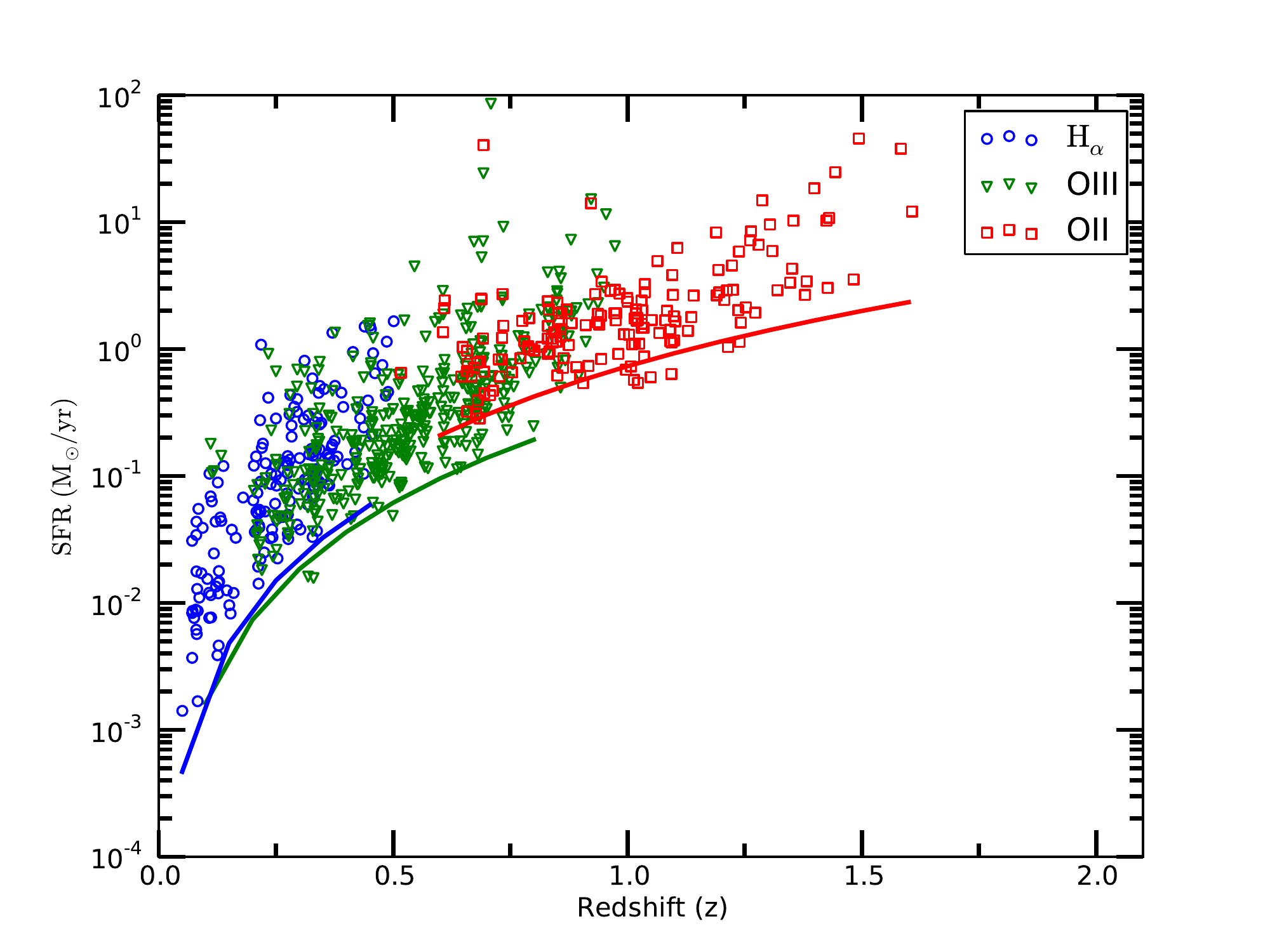}  
\caption{The SFR as a function of redshift for the PEARS  \OIII, \OIII, and \Ha\ ELRs 
with a line grade greater than 2.5. The solid  lines show the SFR corresponding to a
flux limit of ${\rm 1 \times 10^{-17}\ erg/s/cm^2}$, below which 
most line-emitting sources would not be detected (see Figure \ref{correction})
\label{sfrz}}
\end{figure}

\begin{figure} 
\includegraphics[width =7in]{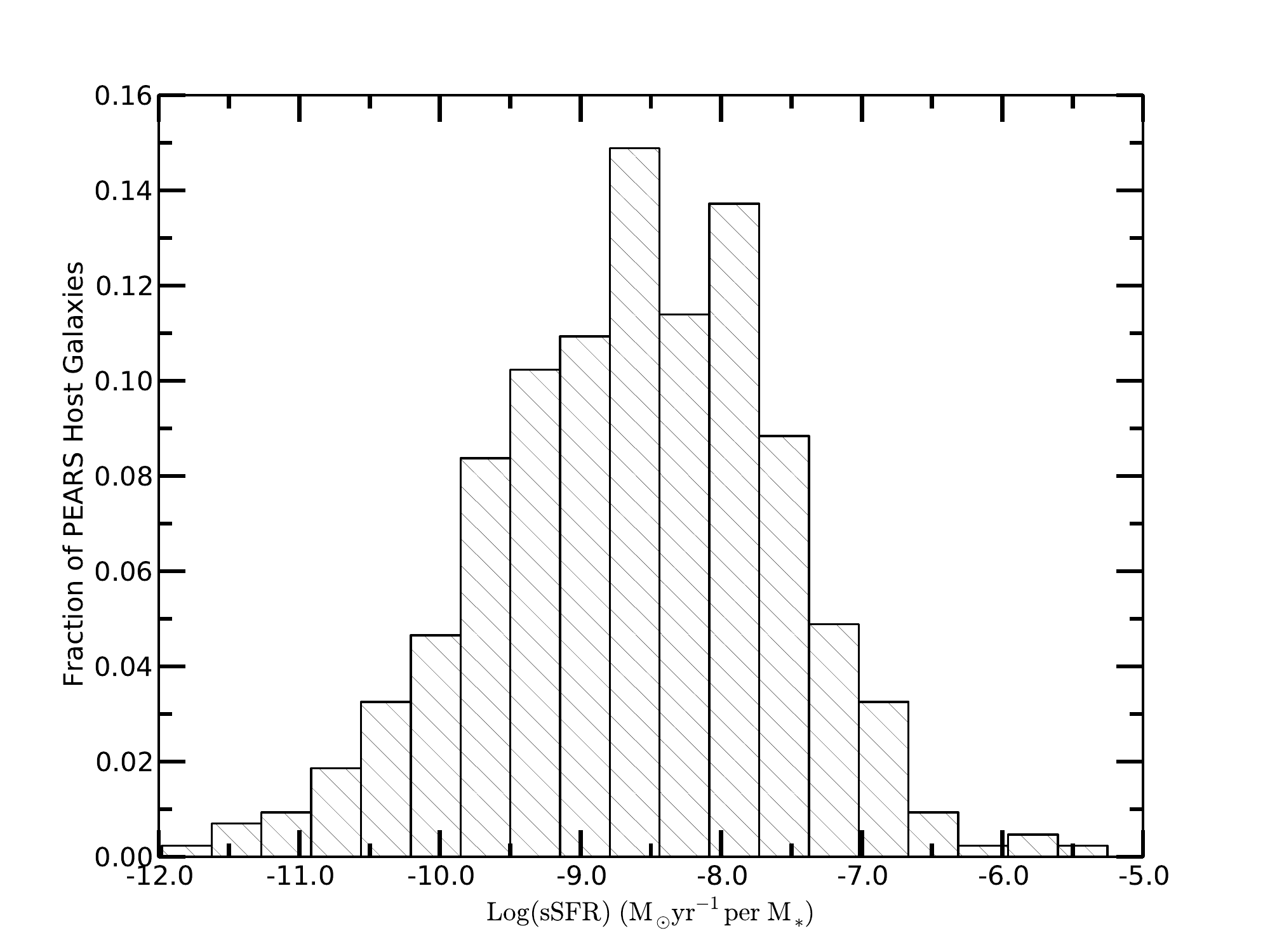} 
\caption{Distribution of the specific SFR (sSFR) for the PEARS host galaxies. The sSFR shown were corrected for extinction using the luminosity dependent dust correction from \citet{hopkins2001}.
\label{sSFRhist}}
\end{figure}

\begin{figure} 
\includegraphics[width =7in]{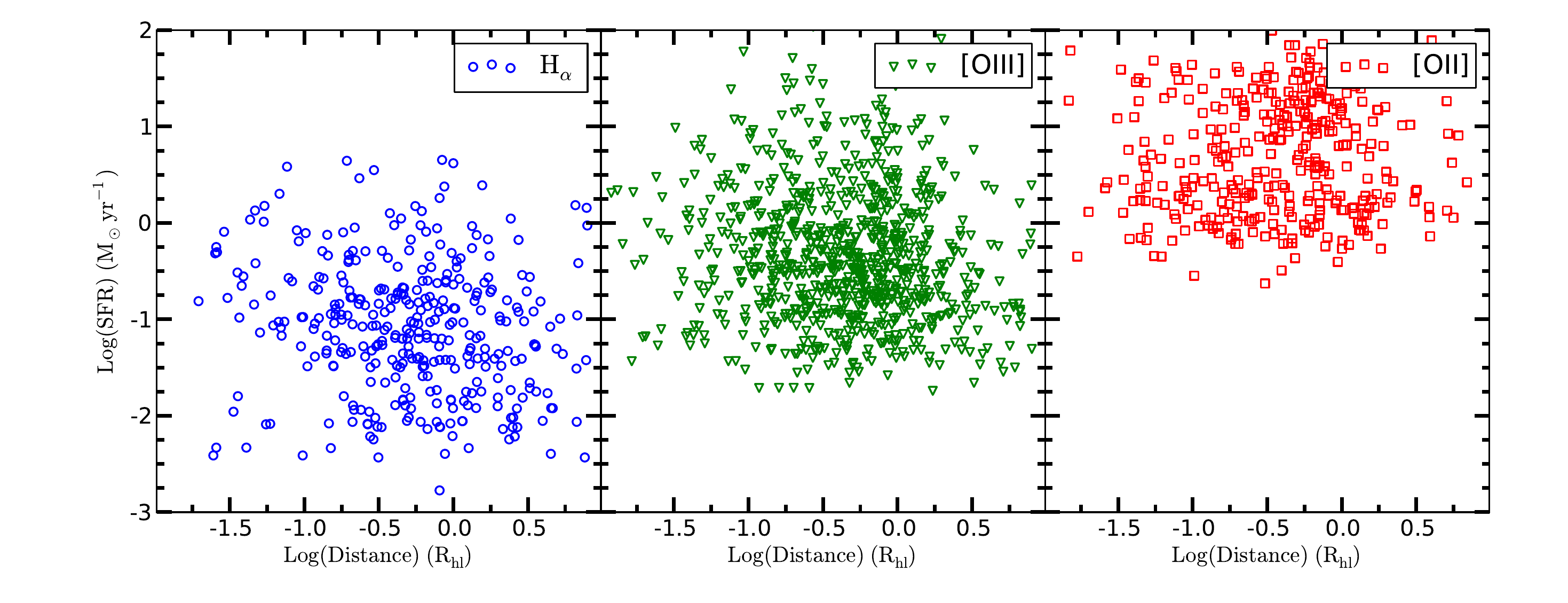}  
\caption{The dust corrected SFR of the PEARS emission line regions plotted as a function of their radial 
position in their host galaxy, normalized to the half light radius of the host galaxy, $R_{hl}$). 
The amount of star-formation appears uncorrelated with the location of the ELR 
in the host galaxy with Pearsons correlation coefficients of -0.03, -0.03 and 0.01 for 
\Ha, \OIII\ and \OII, respectively.  This indicates no correlation exists between the parameters plotted.
These three panels are proxies for the redshift bins of $0 < z < 0.5$, $0.1 < z < 0.9$, and $0.5 < z < 1.5$, respectively.
\label{R2SFR}}
\end{figure}

\begin{figure} 
\includegraphics[width =7in]{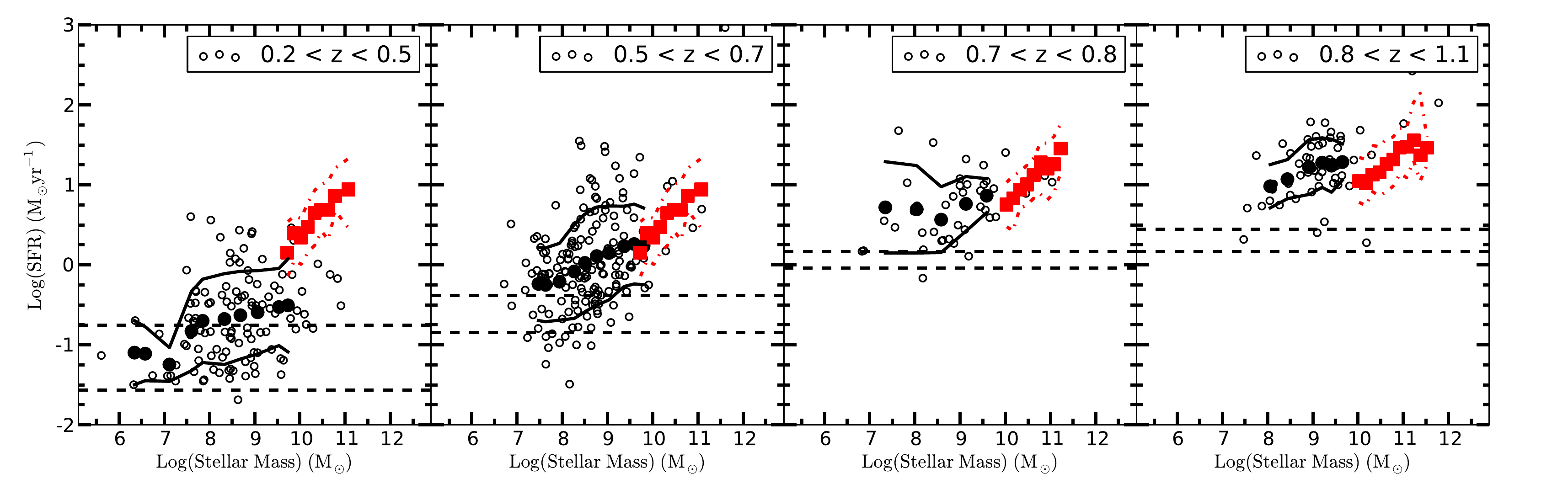} 
\caption{Comparison of the dust corrected SFR with Stellar Mass for the ELG sample (open black circles).  
The data have been binned into 4 redshift ranges for a 1:1 comparison with higher mass 
star-forming galaxies from the Extended Groth Strip \citep{Noeske2007}.  The dash-dotted
red line represents the $\pm$1$\sigma$ of the median values of \citet{Noeske2007}.  The solid black circles 
represents the median for the ELG sample.  The solid black lines represent the 
$\pm$1$\sigma$ of the median values.  The dashed horizontal line represents the 80\% 
completion levels at the minimum and maximum redshifts considered in each panel.
\label{kaicomparison}}
\end{figure}

\indent Assuming that ELGs are representative of star-forming galaxies in general, the depth
of the PEARS-2D study and  large contiguous redshift range allows us to address the relationship between SFR and redshift.
  The SFR was calculated for the \OII\ and \Ha\ emission lines using the 
\citet{kennicutt1998} relations. For \OIII\ lines, which are likely to 
be blended \OIII\ and \Hb\ lines, the relation from Equation 5 in \citet{drozdovsky2005}
was adopted to correct for contamination.  We note that while  \OIII\ emitters cover the broadest redshift range in our sample, \OIII\ is the least 
reliable SFR indicator, especially since we unable to account for the metallicity of each source. The results are shown for individual ELR as a function of redshift in Figure 
\ref{sfrz}.  Also plotted (as solid lines) is the SFR for emission lines with an observed flux of 
${\rm 1 \times 10^{-17} erg/s/cm^2}$. This illustrates our ability to detect emission-lines uniformly from $0<z<1.5$.\\
\indent To analyze the growth of stellar mass in galaxies, it is useful to normalize the computed total galaxy SFRs -- summing up contributions from different ELR when necessary -- by the
 galaxies' stellar masses computed in Section  \ref{hostSED}. The resulting specific SFR (sSFR) allows a comparison
  of all  galaxies' SF activity, in units of the time it would take to build the current stellar mass at their current SFR.
 A histogram of the dust corrected sSFR for the PEARS-2D ELGs is shown in 
Figure \ref{sSFRhist}. The dust corrections were derived using the luminosity dependent dust extinction from \citet{hopkins2001} discussed in Section \ref{hostSED}.
Because the dust corrected sSFR value of a galaxy can be considered as one build-up over the life time of a galaxy --- assuming 
constant star-formation rate --- the PEARS ELGs sSFRs imply a possible stellar mass 
built-up time of a few billions years.
Note, however, that these sSFR estimates should be considered lower limits since  some non-detected star 
formation might be present in the PEARS ELGs.\\
\indent There has been some discussion \citep[e.g.,][]{Guo2011} as to whether the SFR in star-forming 
regions of galaxies should be spatially correlated with the star-forming regions 
within the galaxy. We investigate this possible relation using the PEARS ELG sample. 
Figure \ref{R2SFR} shows a plot of the estimated SFR of each ELR in the PEARS-2D sample, 
separating \Ha, \OIII\ and \OII\ emitting regions as a function of radial distance of 
the ELR (normalized to the continuum half light radius of the galaxy). As this Figure shows, there is no indication for trends as a function of ELR-location for either one of the three types 
of ELGs examined. A simple Pearsons linear correlation test for 
\Ha, \OIII\ and \OII\ yields values of --0.03, --0.03 and --0.01, respectively, indicating 
no statistical correlation between the location of ELRs and SFR from those ELRs.\\
\indent Finally, Figure \ref{kaicomparison} compares dust-corrected SFR against $\it M_{*}$
for the ELGs in the PEARS-2D sample (open circles in all panels).  The ELGs
are plotted in four redshift bins to match the results of \citet{noeske2007}. In that work,
\citet{noeske2007} derived a ``main-sequence'' of star-forming galaxies for field
galaxies in the Extended Groth Strip, complete to Log {\it M} $\sim$ 10.8 
(Figure 1 in that paper).  The red squares in Figure \ref{kaicomparison}
are the median values for the galaxies of \citet{noeske2007}, along with the 
$\pm$1$\sigma$ (dotted red line).  Their conclusion was that there exists a gradual
decline in SF of most galaxies since z$\sim$1.  The implication is that 
the same physics that regulates SF in local disk galaxies is occurring at $z\approx1$, which could
be either due to an evolution in the gas supply or changes in the SF efficiencies.
\citet{noeske2007b} suggested that the slope of their star-formation ``main-sequence'' is related 
to the gas-exhaustion of galaxies, and is related to the age of the galaxy
and its star-formation timescale, all of which are dependent on the galaxy mass.  The PEARS-2D
sample probes galaxies to much lower masses than those in \citet{noeske2007}.
The PEARS galaxies are compared to those in in \citet{noeske2007} 
in Figure \ref{kaicomparison}. 
Our results suggest that the ``main sequence'' , previously found for more massive galaxies, 
exists down to very low stellar masses, $\sim 10^8$ M$_{\sun}$, out to $z\sim 1$.
There is also potentially a flattening of the sSFR versus mass relation for lower mass 
objects (i.e., below $\approx 10^8M_\sun$).
The slope of the "main sequence" can differ for different SFR indicators. However, such a flattening in SFR
 vs. stellar mass, if real, would indicate an even steeper increase of sSFR with decreasing stellar 
 mass than \citet{Noeske2007b} had found at higher masses, and aggravate
  the requirement for a late onset of efficient SF in low mass galaxies, discussed by these authors.
 The dashed horizontal lines in Figure \ref{kaicomparison} show our sensitivity limits at the lower and
higher ends of the redshift ranges shown. It is clear that the flattening of this relation is
 not caused by incompleteness, especially at the higher redshifts.

\subsection{The Host Galaxies}\label{hosts}
\subsubsection{Morphologies}
\begin{figure} 
\includegraphics[width =6.in]{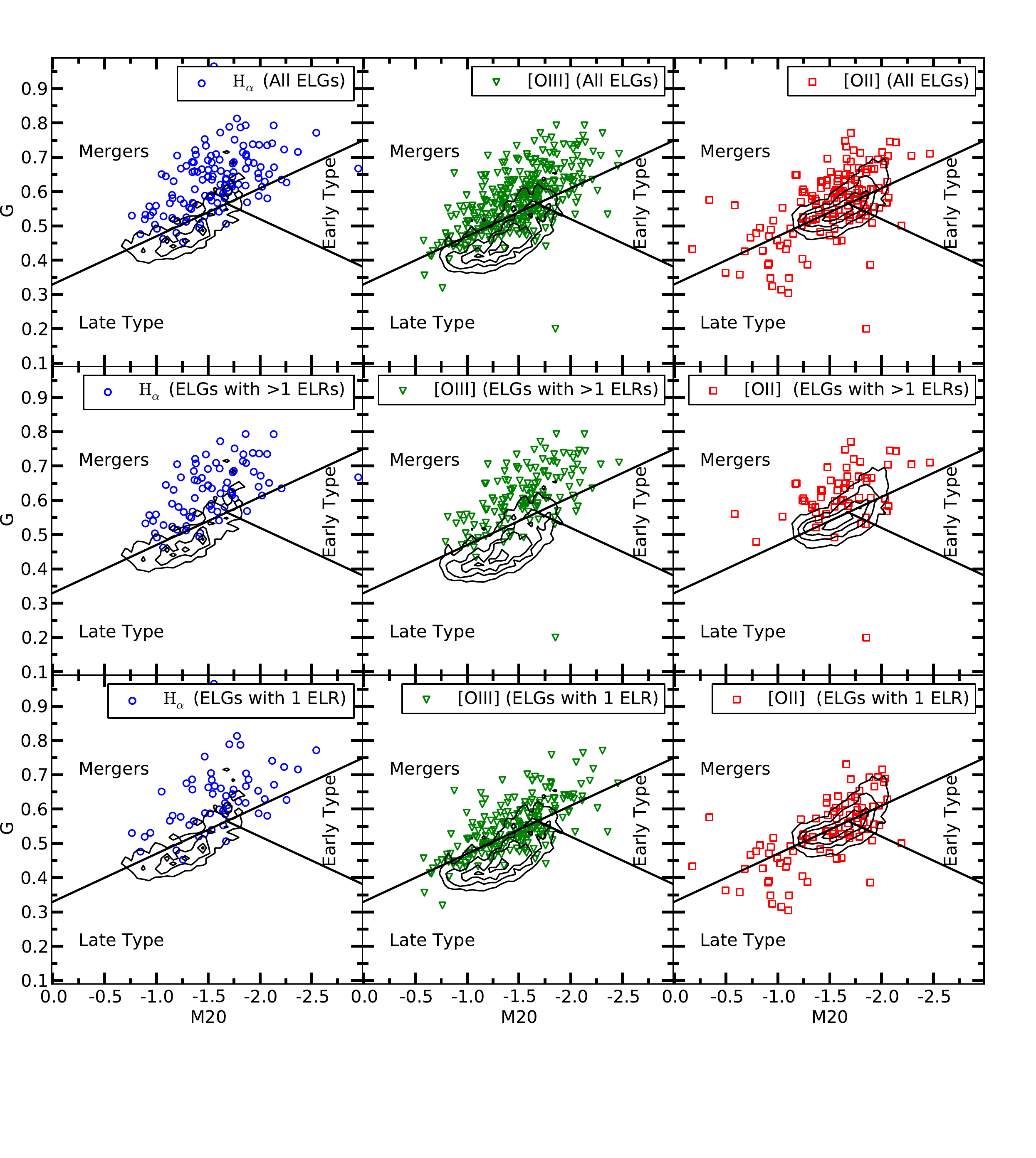}  
\caption{Morphology of the PEARS ELGs as parametrized by the Gini and M20 coefficients at 
the rest-frame wavelength of $4350$\AA. Top row:  PEARS ELGs with \OII, \OIII, and \Ha\  ELRs (red squares, green triangles and blue circles, in the left to right columns, respectively).  Middle row: PEARS ELGs containing multiple ELRs. Bottom row:  PEARS ELGs containing only one ELR. In every panel we  show the rest-frame morphology of the GOODS field galaxies are shown (using contours) with photometric redshift estimates ranges of $0 < z < 0.5$, $0.1 < z < 0.9$, and $0.5 < z < 1.5$ (left to right columns), respectively.
The galaxy hosts of the emission lines that we detected are nearly all above the line (shown in black) 
that separates ''normal'' galaxies (below the line) and ''merging'' galaxies in the nearby 
Universe and is taken from  \citet{Lotz2004}. Most PEARS ELGs are clumpy and have ``merger-like'' \Gini-\M20\ values when observed in the rest-frame wavelength of $4350$\AA.
\label{ginim20}}
\end{figure}

\begin{figure} 
\includegraphics[width =7in]{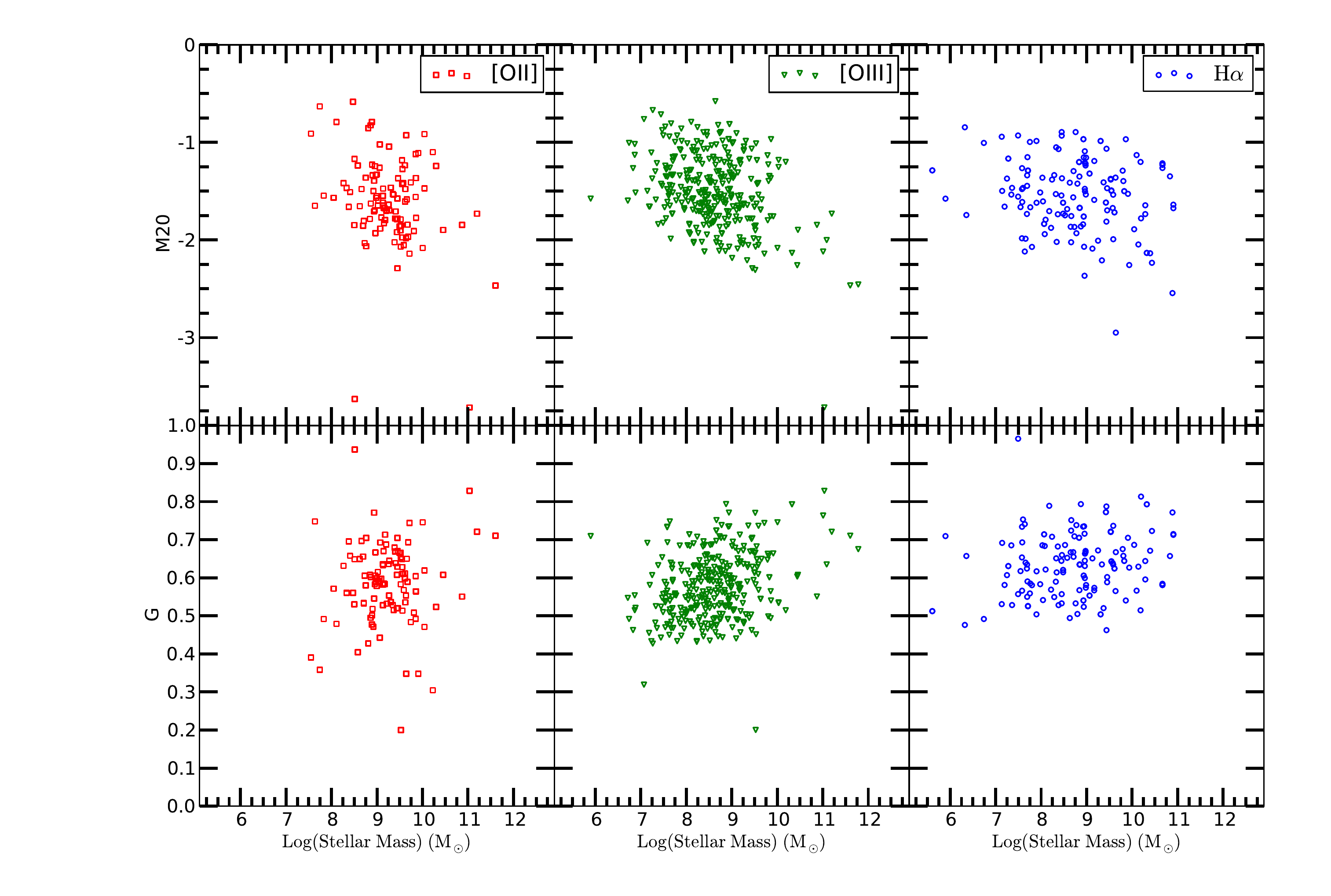} 
\caption{Gini coefficient values of the PEARS host galaxies versus their stellar masses, as estimated from SED fitting. The M20 and the Gini values are shown in the top and bottom row, respectively. The \OII, \OIII\ and \Ha\ host galaxies are shown separately in the left, middle and right most column respectively. There is little evidence for a strong trend between stellar mass and either the M20 or Gini coefficients in our PEARS emission-line host galaxies, as indicated by Pearsons correlation coefficient values of at most  $\approx 0.16$. However, a  mild decrease in M20, and an increase in the Gini coefficient as stellar mass increases can be seen for the \OIII\ and \Ha\ host galaxies (at redshifts of $0.1<z<0.9$ and $0<z<0.5$ respectively).
\label{GM20mass}}
\end{figure}

\indent The PEARS-2D galaxies comprise a remarkably robust sample to test the 
evolution of ELGs and compare their morphologies with physical properties such as 
SFRs and stellar masses.   Unlike many morphological studies, our sample was not 
pre-selected by redshift or luminosities. The PEARS sample was found to be mostly unbiased by the actual morphology of
the host galaxies. It should  be noted, however, that the PEARS sample is dependent
on the strength of emission-lines and would therefore tends to favor the inclusion of low extinction and low metallicity galaxies.
 In this section, we parameterize the morphologies of the host galaxies using the Gini Coefficient 
\Gini\ and \M20\ parameters \citep{Lotz2004}.  The \Gini\ and \M20\ parameters can be 
thought of as proxies for clumpiness and concentration coefficients, respectively, and  have been shown 
to be a good way to distinguish between ''normal'' galaxies and galaxy mergers in the 
local Universe  \citep{Lotz2004} in the blue using the Sloan 
{\it B}$_{\rm J}$, Thuan-Gunn {\it g}, and {\it B}-bands to classify local normal galaxies and rest-frame R-band for all types of mergers.
 Local spiral and 
elliptical galaxies follow a well defined  \Gini -\M20\ sequence  \citep[e.g. Figure 9 in][]{Lotz2004}), 
while mergers have larger \Gini\ and smaller \M20\ values 
\citep{Lotz2004,lotz2008,lotz2010}. In order to compare the PEARS ELGs to galaxies 
in general, we  computed the rest-frame \Gini\ and \M20\ coefficients for both the 
PEARS ELGs as well as the entire GOODS catalog using the public GOODS V2.0 ACS  data. 
These values were measured in all available observed wavelengths, and a rest-frame
$\lambda$ $\sim$ B-band (or $\sim$ 4350 \AA) value was obtained by linearly 
fitting these measurements.\\
\indent As a comparison, field galaxies from  the same GOODS fields are included using the public  GOODS V2.0 ACS 
data.  For these objects, photometric redshifts were used to  derive rest frame B-band values for \Gini\ and \M20.  The rest-frame values of \Gini\ and 
\M20\ were computed by linearly fitting the values measured in each of the available bands.
 The galaxies are all plotted in Figure \ref{ginim20}, which is divided
into three rows of three panels for clarity.  The field galaxies from GOODS are plotted as contours, 
and in each panel the  \Ha\, \OIII\,, and \OII\ ELGs are shown separately.
The solid lines in Figure \ref{ginim20} delineates disturbed galaxies (above the line) 
from ''normal'' galaxies (below the line),  following to  \citet{Lotz2004}.  Also shownm are the regions
containing early-type and late-type ''normal'' galaxies.
When compared to the rest of the GOODS filed galaxies (black contours), the PEARS ELGs  
clearly have higher \Gini\ and \M20\ values, and fall above the fiducial line 
separating quiescent galaxies from active galaxies  \citep[following][]{Lotz2004}. 
This strongly suggests that the PEARS ELGs have perturbed morphologies. 
While it is possible that some, most, or even all of the PEARS host galaxies are ongoing mergers, 
it is difficult to quantify without similar \Gini\ and \M20\ measurements of mergers
using restframe B-band imaging.  We note that in \citet{Lotz2004,lotz2008}, the mergers were 
predominantly Ultraluminous Infrared Galaxies (ULIRGs) in various stages of interaction (i.e.
ranging  from two discernible progenitors to single objects with coalesced nuclei) and 
were based on observations obtained with the {\it F814W} filter using WFPC2 on {\it HST}.  
The median redshift of this heterogeneous sample placed the observations at $\sim$ rest-frame 
{\it R}-band.  As noted in \citet{Taylor-Mager2007,Rothberg2010,Rothberg2013}
the structure and morphology of mergers change as a function of wavelength from UV to near-IR.
Moreover, the stage of the merger can affect concentration indices \citep{Taylor-Mager2007}.
At best, we can state that the PEARS sample is dominated by clumpy systems dis-similar 
to nearby normal galaxies at rest-frame {\it B}-band and {\it may} indicate some type of
merger activity is occuring for some of the sample.\\
\indent  There is an indication that ELGs 
with more than one detected ELR tend to have more disrupted morphologies, while ELGs with a single ELR tend to lay closer 
the the line separating normal galaxies and mergers (bottom three panels), as defined by \citet{Lotz2004}. 
However, there is no correlation between the Gini-M20 values and their computed SFRs and stellar masses.
To test for any correlation correlation, the Pearson Correlation coefficient ({\it r}) 
was used.  It tests the degree of linear correlation between two independent data sets. 
Here, {\it r} ranges in value from --1 to +1 (perfect negative or anti-correlation to perfect 
positive correlation).  The most correlated relation we find is that of the \OIII\ versus stellar mass, shown in Figure \ref{GM20mass},
which which shows a very weak with a value of  {\it r} $=0.16$. All other relations show no statistically significant correlation.

\subsubsection{4350\AA\ Rest-frame Luminosity of ELGs}
\begin{figure} 
\includegraphics[width =7in]{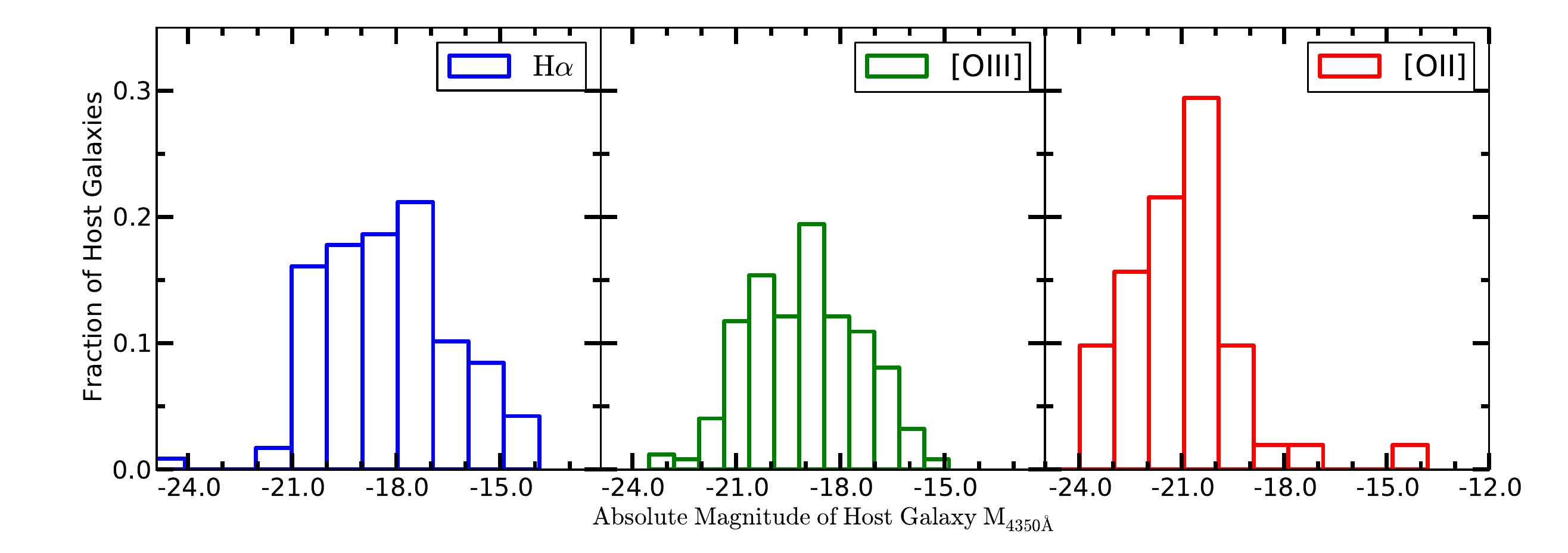} 
\caption{Distributions of the $4350$\ \AA\ rest-frame absolute magnitude of the host galaxies 
of the PEARS \Ha, \OIII\ and \OII\ emission line (redshift ranges of  $0 < z < 0.5$, $0.1 < z < 0.9$, and $0.5 < z < 1.5$), respectively.
\label{Mhist}}
\end{figure}

\begin{figure} 
\includegraphics[width =7in]{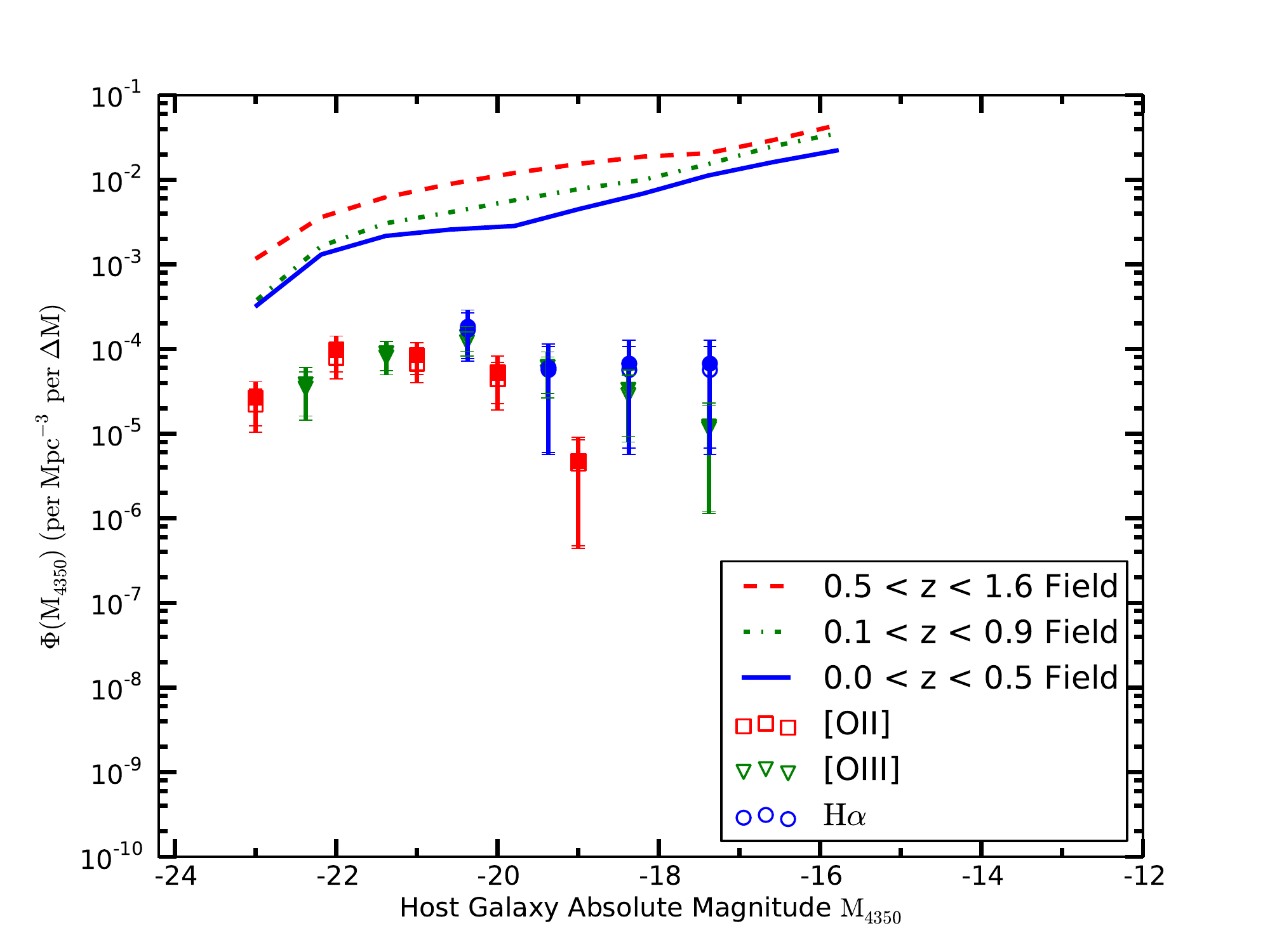} 
\caption{Rest-frame $4350$\AA\ luminosity functions for the PEARS host 
galaxies of  PEARS \Ha, \OIII\ and \OII\ emission lines  (symbols with error bars,  for the redshift ranges of $0 < z < 0.5$, $0.1 < z < 0.9$, and $0.5 < z < 1.5$, respectively).  Only galaxies with at least one PEARS emission line with with a SFR $> 6\ M_\sun yr^{-1}$ are shown (which corresponds to an emission line at $z=1.5$\ with an observed flux of  ${\rm  3 \times 10^{-17}\ erg/s/cm^2}$).
Both the completeness corrected (filled symbols) and uncorrected (open symbols) density 
estimates are presented. Although the GOODS data are more than deep enough to allow us to detect host galaxies with $M<-18$\ mag at all redshifts (solid curves), we detect no galaxy with $M>-18$\ mag at $0.5 < z < 1.6$ with \OII\ emission.
\label{Mlum}}
\end{figure}

\indent The underlying host galaxy luminosity may provide additional information
about the nature of the ELGs, and how they compare to other galaxies in the field.
Rest-frame absolute magnitudes at 4350\AA\ ({\it M}$_{\rm 4350}$) were computed for both
the ELGs and the GOODS field galaxies.  Figure \ref{Mhist} shows a histogram
distribution of  {\it M}$_{\rm 4350}$ for the ELGs, divided into three panels ---
one for each emission line.
The median  {\it M}$_{\rm 4350}$\ values for each emission line are:
 --21.2 mag for \OII; --19.0 mag for \OIII; and --18.2 mag for \Ha. \\
\indent One important question is whether the ELGs are representative of other
galaxies within the same volume.  As discussed in the Introduction, ELGs are
very useful for probing the evolution of the SFR, not only out to more distant
epochs, but also  to fainter luminosities (and thus lower masses) than other galaxies.
Figure \ref{Mlum} compares the luminosity function of the ELGs --- separated by
emission line which sample different redshift bins ---
that form stars strongly enough to be detected at all redshifts in the PEARS sample. 
Our survey line sensitivity limit of ${\rm  3 \times 10^{-17}\ erg/s/cm^2}$ corresponds to an SFR of $6\ M_\sun yr^{-1}$,
at the maximum redshift of $z\sim 1.5$\ that is
probed by PEARS. This SFR is in between the values observed in a strong star forming galaxy such as M51 (NGC 5194), with an SFR of $\sim 3.5\ M_\sun yr^{-1}$\ \citep{Calzetti2005}, and the SFR of starburst galaxies such as M82 with an SFR of $10-30\ M_\sun yr^{-1}$\ \citep{Beswick2006}. 
In Figure \ref{Mlum}, 
the line luminosity function is plotted for these star forming galaxies and compared to the the luminosity 
function of the  GOODS field galaxies. While the luminosity functions 
of the GOODS field galaxies increase monotonically in a power law manner, 
the volume density of strong line-emitters decreases quickly in the redshift bin $0.5<z<1.5$ for host galaxies with $M>-20$\ mag.
The GOODS data are faint enough to detect host galaxies that are much fainter  ($M \sim -15$\ mag). The PEARS
survey is sensitive enough to detect the emission lines from such strong star forming galaxies, but we did not identify emission lines 
galaxies with $M > -18$\ mag in the higher redshift range of the survey. 
It is unlikely that this apparent decrease is caused by incompleteness because the sample is restricted to include only objects with line fluxes that are comfortably above 
the incompleteness limit and, as long as a galaxy forms stars at a rate greater than $6 {\rm M_\sun\ yr^{-1}}$, this 
will result in an emission line that is bright enough to be detected and identified by the PEARS-2D method. The PEARS survey might
simply be missing galaxies at higher redshifts if these objects are intrinsically more dusty at higher redshift, but the required 
amount of extinction are large and we see not evidence for an increase in the SED derived extinction for the objects we detect, as a function 
of redshift.

\begin{figure} 
\includegraphics[width =7in]{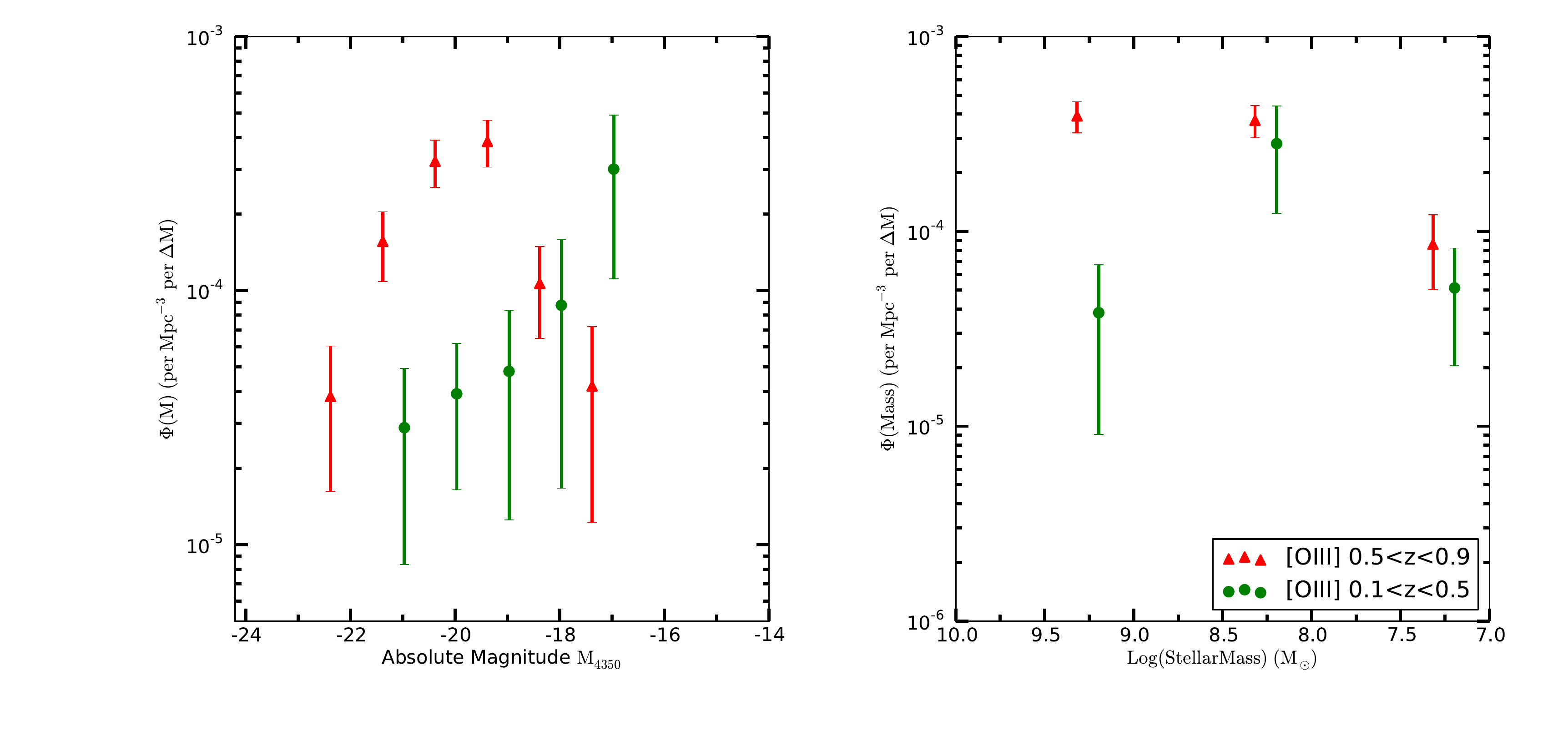} 
\caption{Left Panel: Rest-frame  $4350$\AA\ luminosity function
for the host GOODS galaxies, where at least one \OIII\ emission-line region was detected 
with a ${\rm SFR> 1.7\ M_\sun\ yr^{-1}}$ (corresponding to our 
flux limit at the maximum observable redshift of 0.9). We plot the densities of host 
galaxies in the lower redshift range of ${\rm 0.1<z<0.5}$ as green circles and the 
ones at ${\rm 0.5<z<0.9}$ as red triangles. There is a strong 
decrease in the density of faint (${\rm M_{4350} > -19\ mag}$) host galaxies at higher redshifts, 
that does not exist at lower redshifts. Right Panel: The corresponding host galaxy mass 
function for the data shown on the left panel. There is a strong, ten-fold decrease in the 
number density of galaxies with \OIII\ emission with stellar masses greater than 
$10^{9}\ M_\sun$ in the lower redshift bin while the density of \OIII\ emitting galaxies 
remains the same for less massive galaxies.
\label{MassLum2}}
\end{figure}

\indent We can also quantify any redshift dependence of the volume density of line-emitting galaxies by examining the volume densities of just the \OIII\ 
emitting galaxies at redshifts from $z\approx 0.1$\ to $z\approx0.9$.  This is the redshift range with the largest
number of ELGs and \OIII\ lines should be less sensitive to dust than the \OII\ lines discussed above.  When examining
the \OIII\ line emitters, we now restrict our sample to galaxies with $SFR> 1.7 M_\sun yr^{-1}$, 
our SFR completeness limit at $z=0.9$, which is smaller than the limit we used when we included the \OII\ ELGs, but
still select galaxies with robust star formation -- while maintaining a sample size that is as large as possible.  The \OIII\ host galaxy sample was  divided into two distinct redshift ranges 
(0.1 $<$ z $<$ 0.5 and 0.5 $<$ z $<$ 0.9) and the results
are plotted in Figure \ref{MassLum2}. This Figure shows the luminosity functions as a function of both  {\it M}$_{\rm 4350}$ ({\it left}) 
and Stellar Mass ({\it right}).  The left panel of Figure \ref{MassLum2} confirms that 
there appears to be a relatively small number of faint galaxies with detected \OIII\ 
emission at higher redshifts.  Recall that these host galaxies were selected solely 
based on the direct and independent detection of \OIII\ in emission, and thus were selected independently
of their observed size and host luminosity. \\
\indent In the right panel of Figure \ref{MassLum2}, the stellar masses are compared
for the two redshift ranges (same limits on sample selection as in the left panel).  The
stellar mass distribution of galaxies with detected \OIII\ emission differs significantly.
At lower redshift there appears to be fewer massive galaxies with detected star-formation $>1.7 M_\sun yr^{-1}$. 
We conclude that the ratio of star-forming massive galaxies to passive massive galaxies is higher at high redshift, 
which is a result that is  consistent with downsizing \citep[e.g.,][]{cowie1996}.

\section{Conclusions}
\indent We have presented a sample of ELGs selected independently by their emission-lines
without {\it a priori} knowledge of their host galaxies properties.  The methodology used
(PEARS-2D) is based on {\it direct} detection of emissions-line from {\it HST} 
slitless grism spectroscopy, with the added bonus of being able to detect multiple
ELRs within a single galaxy.  This  has yielded  a sample, which 
is effectively random and blind to other parameters.  Using the wealth of ancillary
data, the properties of the underlying host galaxies were compared
with the SFR histories derived from the ELRs.  The key results are summarized below:\\
\noindent 1)  There is evidence for evolution in the luminosity functions of  the 
\Ha,\OIII\ and \OII\ emission lines. The luminosity function slopes flatten as a function of redshift.\\
\noindent 2)The morphology of the host galaxies clearly indicates that these objects
 are clumpy, although we detect no correlation between their morphology 
and our stellar mass estimates or star-formation intensity (sSFR).\\
\noindent 3)The mass-density function of \OIII\ emitting galaxies at ${\rm 0<z<0.9}$ 
strongly decreases with redshift. The number density of objects with stellar masses greater that 
${\rm \sim 10^{10}\ M_{\sun}}$ undergoing strong star-formation decreases at lower redshifts. 
This supports the idea of galaxy downsizing \citep[][e.g.]{cowie1996}.\\
\noindent  The results presented here also demonstrate the clear advantage of 
space-based grism spectroscopy using multiple position angles.  Such observations are able to probe deeper than
similar ground-based studies.  The PEARS-2D method also provides a method
for detecting multiple ELRs, and allows spatial information about SF to be derived for
galaxies.  Future work will include using the WFC3 near-IR grism mode, with observed
wavelength coverage of 0.8-1.6$\micron$.  This will allow us to probe to significantly
higher redshifts, and determine whether the trends reported here continue to earlier 
epochs. After 2018, this work can also be done at much higher sensitivities with the JWST FGS grism at $1-2 \mu m$, and the JWST NIRcam prisms at $2.5-5 \mu m$.

\acknowledgments
{\em Acknowledgments} - NP wishes to thank F. Pierfederici for his help during the preparation of this manuscript.
This research made use of the OSX version of SCISOFT assembled by Dr. Nor Pirzkal and F. Pierfederici. 
\clearpage
\begin{deluxetable}{cccccccccccccccccc}
\tabletypesize{\footnotesize}
\setlength{\tabcolsep}{0.06in}
\tablewidth{0pt}

\tablecaption{Summary of emission lines detected in the PEARS survey using the PEARS-2D method.\label{table1}}
\tablehead{
\colhead{PEARS} &
\colhead{R.A.}  &
\colhead{Dec.}  &
\colhead{No.} &
\colhead{Exposure} &
\colhead{ No.} &
\colhead{ No.} &
\multicolumn{10}{c}{No. of lines} \\
\cline{8-17}

\colhead{Field} &
\colhead{} &
\colhead{} &
\colhead{PA\tablenotemark{a}} &
\colhead{(s)\tablenotemark{b}} &
\colhead{Objects} &

\colhead{Knots}&  
\colhead{\OII}&
\colhead{\OIII}&
\colhead{\Ha}&
\colhead{\CIV}&
\colhead{\CIII}&
\colhead{\MgII}&
\colhead{\NeIII}&
\colhead{\Hg}&
\colhead{\Lya}&
\colhead{noID}&
\colhead{All\tablenotemark{c}}
}
\rotate
\startdata
PEARS-N-1 &189.1852503 & +62.2032822 & 3 & 44708 & 153 & 167 & 48 & 73 & 26 & 5 & 6 & 6 & 1 & 5 & 2 & 1 & 173 \\
PEARS-N-2 &189.1877163 & +62.2548588& 3 & 44252  & 90 & 96 & 29 & 50 & 15 & 1 & 4 & 3 & 4 & 4 & 1 & 5 & 116 \\
PEARS-N-3 &189.3100669 & +62.2924237& 3 & 44708 & 98 & 104 & 33 & 43 & 16 & 5 & 9 & 5 & 1 & 2 & 1 & 2 & 117 \\
PEARS-N-4 &189.3720309 & +62.3201389& 3 & 44708 & 91 & 98 & 23 & 42 & 35 & 1 & 10 & 3 & 0 & 2 & 1 & 6 & 123 \\
PEARS-S-HUDF &53.16231255 &-27.7911063& 4 & 89819 & 152 & 166 & 46 & 61 & 40 & 7 & 7 & 13 & 0 & 3 & 2 & 14 & 193 \\
PEARS-S-1 &53.16967450 & --27.9014641& 3 & 43733 & 54 & 61 & 10 & 34 & 28 & 0 & 0 & 0 & 0 & 2 & 0 & 2 & 76 \\
PEARS-S-2 &53.17745315 & --27.8416506& 4 & 51583 & 52 & 54 & 20 & 40 & 5 & 0 & 0 & 0 & 1 & 2 & 0 & 2 & 70 \\
PEARS-S-3 &53.11987485 & --27.7396665& 3 & 44186 & 112 & 127 & 29 & 59 & 34 & 0 & 6 & 5 & 1 & 3 & 1 & 7 & 145 \\
PEARS-S-4 &53.06664343 &--27.7088154& 4 & 44084 & 123 & 131 & 31 & 62 & 20 & 3 & 4 & 9 & 2 & 6 & 3 & 9 & 149 \\
\cline{1-17}
PEARS-N\tablenotemark{d} & & & & & 417 & 451 & 133 & 208 & 92 & 12 & 29 & 17 & 6 & 13 & 5 & 14 & 529 \\
PEARS-S\tablenotemark{e} & & & & & 489 & 535 & 136 & 256 & 127 & 10 & 17 & 27 & 4 & 16 & 6 & 34 & 633 \\
PEARS TOTAL\tablenotemark{f}& & &  & & 906 &  986 & 269 & 464 &  219 & 22 & 46 & 44 & 10 & 29 & 11 & 48 & 1162\\

\enddata
\tablecomments{(a) Number of HST orientations at which this field was observed.\\
(b) Total exposure time (in s) of all the data obtained for this field, including all orientations.\\
(c) Includes  \Lya, \CIV, \CIII, \MgII, \OII, \NeIII, \OIII, \Hg, and \Ha.\\
(d) Sum for all the PEARS-N fields.\\
(e) Sum for all the PEARS-S fields.\\
(g) Sum for all of the PEARS fields.
}
\end{deluxetable}

\input{table2s
.tex}

\begin{deluxetable}{cccccccccccc}
\tablecaption{Luminosity Function}\label{lumfcttable}
\tabletypesize{\footnotesize}
\setlength{\tabcolsep}{0.06in}
\tablewidth{0pt}
\tablehead{
\colhead{Method} &
\colhead{Line} &
\colhead{Redshift} &
\multicolumn{3}{c}{PEARS-N} &
\multicolumn{3}{c}{PEARS-S} &
\multicolumn{3}{c}{PEARS} \\
\cline{4-6}
\cline{7-9}
\cline{10-12}
\colhead{} &
\colhead{} &
\colhead{Range} &
\colhead{${\rm L_*}$} &
\colhead{$\alpha$} &
\colhead{$\Phi_*$} &
\colhead{${\rm L_*}$} &
\colhead{$\alpha$} &
\colhead{$\Phi_*$} &
\colhead{${\rm L_*}$} &
\colhead{$\alpha$} &
\colhead{$\Phi_*$}
}
\rotate
\startdata
& OII & $0.5 < z < 1.6$ & $41.45_{-0.14}^{+1.56}$ & $-1.23_{-0.87}^{+0.42}$ & $-4.08_{-5.83}^{+0.45}$  & $41.95_{-0.48}^{+0.48}$ & $-1.49_{-0.35}^{+0.53}$ & $-4.88_{-1.27}^{+0.98}$  & $41.75_{-0.29}^{+0.57}$ & $-1.44_{-0.42}^{+0.38}$ & $-4.63_{-4.66}^{+0.52}$ \\

STY & OIII & $0.1 < z < 0.9$ & $41.65_{-0.12}^{+0.57}$ & $-1.23_{-0.21}^{+0.14}$ & $-3.45_{-0.73}^{+0.30}$  & $41.68_{0.00}^{+0.36}$ & $-1.19_{-0.23}^{+0.10}$ & $-3.44_{-0.66}^{+0.19}$  & $41.67_{-0.13}^{+0.10}$ & $-1.21_{-0.12}^{+0.11}$ & $-3.45_{-0.42}^{+0.21}$ \\

& \Ha & $0.0 < z < 0.5$ & $41.01_{-0.22}^{+0.43}$ & $-1.14_{-0.29}^{+0.26}$ & $-2.99_{-1.52}^{+0.39}$  & $40.83_{-0.24}^{+0.01}$ & $-0.86_{-0.18}^{+0.29}$ & $-2.50_{-0.27}^{+0.32}$  & $40.90_{-0.03}^{+0.07}$ & $-0.97_{-0.19}^{+0.11}$ & $-2.72_{-0.43}^{+0.16}$ \\

\hline

& \OII & $0.5 < z < 1.6$ & $44.46_{-0.14}^{+0.54}$ & $-1.84_{-0.15}^{+0.11}$ & $-6.30_{-1.03}^{+0.06}$ & $44.35_{-1.61}^{+0.65}$ & $-1.93_{-0.10}^{+0.12}$ & $-6.62_{-0.94}^{+1.59}$ & $43.24_{-1.07}^{+1.76}$ & $-1.93_{-0.08}^{+0.14}$ & $-5.49_{-1.99}^{+1.24}$ \\ 
\vmax & \OIII & $0.10 < z < 0.90$ & $40.87_{-0.13}^{+0.07}$ & $-0.77_{-0.10}^{+0.23}$ & $-2.17_{-0.12}^{+0.08}$ & $41.45_{-0.17}^{+0.21}$ & $-1.19_{-0.12}^{+0.15}$ & $-2.67_{-0.22}^{+0.15}$ & $41.31_{-0.09}^{+0.09}$ & $-1.21_{-0.07}^{+0.08}$ & $-2.58_{-0.09}^{+0.09}$ \\ 
& \Ha & $0.00 < z < 0.49$ & $40.89_{-0.17}^{+0.13}$ & $-1.10_{-0.11}^{+0.13}$ & $-2.36_{-0.21}^{+0.11}$ & $41.42_{-0.26}^{+0.44}$ & $-1.45_{-0.10}^{+0.06}$ & $-2.89_{-0.44}^{+0.14}$ & $41.01_{-0.09}^{+0.06}$ & $-1.24_{-0.04}^{+0.05}$ & $-2.47_{-0.07}^{+0.07}$ \\ 
\enddata
\end{deluxetable}

\begin{deluxetable}{cccccc}
\tablecaption{Properties of significantly detected (grade$>$2.5) emission lines in the PEARS sample\label{linefluxes}}
\tablehead{
\colhead{Line} &
\colhead{Number} &
\colhead{${ <z>}$} &
\multicolumn{2}{c}{Flux ${\rm (erg/s/cm^2)}$} \\
\cline{4-5}
\colhead{} &
\colhead{Detected} &
\colhead{} &
\colhead{Average} &
\colhead{Median} &

}

\startdata
\Ha & 174 & 0.26 & $9.44 \times 10^{-16}$ & $5.54 \times 10^{-17}$   \\
\OIII & 401 & 0.54 & $9.65 \times 10^{-17}$ & $4.13 \times 10^{-17}$    \\
\OII & 167 & 0.91 & $4.19 \times 10^{-17}$ & $2.49 \times 10^{-17}$    \\
\enddata
\end{deluxetable}

\clearpage

\clearpage

\clearpage

\clearpage

\clearpage

\clearpage

\clearpage

\clearpage

\clearpage

\clearpage

\clearpage

\clearpage

\clearpage

\clearpage

\clearpage

\clearpage

\clearpage

\clearpage

\end{document}